\documentclass[10pt]{article}

\usepackage{graphicx}

\usepackage{ifthen}
\usepackage[dvipsnames]{xcolor}
\usepackage{colortbl}
\usepackage{ifthen}
\usepackage{hyperref}
\usepackage[totoc]{idxlayout}
\usepackage{listings}
\usepackage{makecell}
\renewcommand{\hline}{\Xhline{2\arrayrulewidth}}

\usepackage{datetime}
\usepackage{amsmath}
\usepackage{amsthm}
\usepackage{amsfonts}
\usepackage{amssymb}

\definecolor{myorange}{rgb}{0.843, 0.678, 0.000}
\definecolor{mygreen}{rgb}{0,0.6,0}
\definecolor{mygray}{rgb}{0.5,0.5,0.5}
\definecolor{mymauve}{rgb}{0.58,0,0.82}

\renewcommand{\em}{\it}   

\renewcommand{\arraystretch}{1.4}

\newcolumntype{I}{!{\vrule width 1.5pt}}
\newlength\savedwidth
\newcommand\whline{\noalign{\global\savedwidth\arrayrulewidth
                            \global\arrayrulewidth 1.05pt}%
           \hline
           \noalign{\global\arrayrulewidth\savedwidth}}

\newcommand{\FlaTwoByTwo}[4]{
\left(
\begin{array}{c I c}
#1 & #2 \\ \whline
#3 & #4
\end{array}
\right)
}

\newcommand{\FlaTwoByOne}[2]{
\left(
\begin{array}{c}
#1 \\ \whline
#2
\end{array}
\right)
}

\newcommand{\FlaOneByTwo}[2]{
\left(
\begin{array}{c I c}
#1 & #2
\end{array}
\right)
}

\newcommand{\FlaThreeByThreeTL}[9]{
\left(
\begin{array}{c | c I c}
#1 & #2 & #3 \\ \hline
#4 & #5 & #6 \\ \whline
#7 & #8 & #9
\end{array}
\right)
}

\newcommand{\FlaThreeByThreeBR}[9]{
\left(
\begin{array}{c I c | c}
#1 & #2 & #3 \\ \whline 
#4 & #5 & #6 \\ \hline
#7 & #8 & #9 
\end{array}
\right) 
}

\newcommand{\FlaThreeByOneT}[3]{
\left(
\begin{array}{c}
#1 \\ \hline
#2 \\ \whline
#3
\end{array}
\right)
}

\newcommand{\FlaThreeByOneB}[3]{
\left(
\begin{array}{c}
#1 \\ \whline
#2 \\ \hline
#3
\end{array}
\right)
}

\newcommand{\operation}{}

\newcommand{\routinename}{}

\newcommand{\precondition}{~}

\newcommand{\postcondition}{~}

\newcommand{\invariant}{~}

\newcommand{\guard}{~}

\newcommand{\partitionings}{~}

\newcommand{\partitionsizes}{~}

\newcommand{\blocksize}{blank}

\newcommand{\repartitionings}{~}

\newcommand{\repartitionsizes}{~}

\newcommand{\moveboundaries}{~}

\newcommand{\beforeupdate}{~}

\newcommand{\afterupdate}{~}

\newcommand{\update}{~}

\newcommand{\resetsteps}{

\renewcommand{\operation}{\phantom{[A] = op( A )}}

\renewcommand{\routinename}{\operation}

\renewcommand{\precondition}{\phantom{A = \widehat A}}

\renewcommand{\postcondition}{\phantom{A = \widehat A}}

\renewcommand{\invariant}{\phantom{ \FlaTwoByTwo{A_{TL}}{A_{TR}}{A_{BL}}{A_{BR}} =
		\FlaTwoByTwo{A_{TL}}{A_{TR}}{A_{BL}}{A_{BR}}
		\wedge
		\FlaTwoByTwo{A_{TL}}{A_{TR}}{A_{BL}}{A_{BR}} =
		\FlaTwoByTwo{A_{TL}}{A_{TR}}{A_{BL}}{A_{BR}}~~~~~~~
		}}

\renewcommand{\blocksize}{blank}

\renewcommand{\guard}{\phantom{m( A_{BL} ) < m( A )}}

\renewcommand{\partitionings}{
$
\phantom{\FlaTwoByTwo{A_{TL}}{A_{TR}}{A_{BL}}{A_{BR}}
\rightarrow
\FlaThreeByThreeBR
   {A_{00}}{a_{01}}{A_{02}}
   {a_{10}^T}{\alpha_{11}}{a_{12}^T}
   {A_{20}}{a_{21}}{A_{22}}}   
$
}

\renewcommand{\partitionsizes}{$ \phantom{m( A )} $}

\renewcommand{\repartitionings}{
$
\phantom{\FlaTwoByTwo{A_{TL}}{A_{TR}}{A_{BL}}{A_{BR}}
\rightarrow
\FlaThreeByThreeBR
   {A_{00}}{a_{01}}{A_{02}}
   {a_{10}^T}{\alpha_{11}}{a_{12}^T}
   {A_{20}}{a_{21}}{A_{22}}}   
$
}

\renewcommand{\repartitionsizes}{$\phantom{m(A)}$}

\renewcommand{\moveboundaries}{
$
\phantom{\FlaTwoByTwo{A_{TL}}{A_{TR}}{A_{BL}}{A_{BR}}
\rightarrow
\FlaThreeByThreeBR
   {A_{00}}{a_{01}}{A_{02}}
   {a_{10}^T}{\alpha_{11}}{a_{12}^T}
   {A_{20}}{a_{21}}{A_{22}}}   
$
}

\renewcommand{\beforeupdate}{
\phantom{\FlaTwoByTwo{A_{TL}}{A_{TR}}{A_{BL}}{A_{BR}}
\rightarrow
\FlaThreeByThreeBR
   {A_{00}}{a_{01}}{A_{02}}
   {a_{10}^T}{\alpha_{11}}{a_{12}^T}
   {A_{20}}{a_{21}}{A_{22}}}   
}

\renewcommand{\afterupdate}{
\phantom{\FlaTwoByTwo{A_{TL}}{A_{TR}}{A_{BL}}{A_{BR}}
\rightarrow
\FlaThreeByThreeBR
   {A_{00}}{a_{01}}{A_{02}}
   {a_{10}^T}{\alpha_{11}}{a_{12}^T}
   {A_{20}}{a_{21}}{A_{22}}}   
}

\renewcommand{\update}{
\phantom{$
\begin{array}{l}
\\
\\
\\
\end{array}
$}
}
}

\newcommand{\NoShow}[1]{}

\newlength{\mypaddingskip}
\newcommand{\mypadding}[1][2pt]{\\[\dimexpr #1 + \mypaddingskip]}

\newcommand{\FlaAlgorithm}{
\begin{tabular}{|l|} \hline
$\mbox{\color{blue}Algorithm:~}\routinename$
\\ \hline
\mypadding
\partitionings
\\
$\mbox{\color{blue} ~~~where~}$ \partitionsizes 
\\ 
$\mbox{\color{blue}while~} \ShowGuard \mbox{~\color{blue} do}$
\\
\ifthenelse{\equal{\blocksize}{1}}{}%
{%
\ifthenelse{ \equal{\blocksize}{blank} }{}%
{~~~~{\bf Determine block size $ \blocksize $}\\}%
}
~~~~ 
\repartitionings
\mypadding 
\NoShow{\\
~~~$\mbox{\color{blue} ~~~where~}$ \repartitionsizes
}
\\ \hline

\mypadding
~~~~  \update 

\mypadding
\\ \hline
~~~~ 
\mypadding
\moveboundaries 
\\
$\mbox{\color{blue} endwhile} $
\\ \hline 
\end{tabular}
}

\newcounter{WSStep}
\newcommand{\ShowPrecondition}{\ifthenelse{\value{WSStep}<1}%
   {{\color{white} \precondition}}
   {\ifthenelse{\value{WSStep}=1}%
    {\color{red} \precondition}
    {\color{black} \precondition}}}
\newcommand{\ShowPostcondition}{\ifthenelse{\value{WSStep}<1}%
   {{\color{white} \postcondition}}
   {\ifthenelse{\value{WSStep}=1}%
    {\color{red} \postcondition}
    {\color{black} \postcondition}}}
\newcommand{\ShowInvariant}{\ifthenelse{\value{WSStep}<2}%
   {{\color{white} \invariant}}
   {\ifthenelse{\value{WSStep}=2}%
    {\color{red} \invariant}
    {\color{black} \invariant}}}
\newcommand{\ShowGuard}{\ifthenelse{\value{WSStep}<3}%
   {{\color{lightgray!25} \guard}}
   {\ifthenelse{\value{WSStep}=3}%
    {\color{red} \guard}
    {\color{black} \guard}}}
\newcommand{\ShowGuardTwo}{\ifthenelse{\value{WSStep}<3}%
   {{\color{white} \guard}}
   {\ifthenelse{\value{WSStep}=3}%
    {\color{red} \guard}
    {\color{black} \guard}}}
\newcommand{\ShowPartitionings}{\ifthenelse{\value{WSStep}<4}%
   {{\color{lightgray!25} \partitionings}}%
   {\ifthenelse{\value{WSStep}=4}%
    {\color{red} \partitionings}%
    {\color{black} \partitionings}}}
\newcommand{\ShowPartitionSizes}{\ifthenelse{\value{WSStep}<4}%
   {{\color{lightgray!25} \partitionsizes}}
   {\ifthenelse{\value{WSStep}=4}%
    {\color{red} \partitionsizes}
    {\color{black} \partitionsizes}}}
\newcommand{\ShowRepartitionings}{\ifthenelse{\value{WSStep}<5}%
   {{\color{lightgray!25} \repartitionings}}
   {\ifthenelse{\value{WSStep}=5}%
    {\color{red} \repartitionings}
    {\color{black} \repartitionings}}}
\newcommand{\ShowRepartitionSizes}{\ifthenelse{\value{WSStep}<5}%
   {{\color{lightgray!25} \repartitionsizes}}
   {\ifthenelse{\value{WSStep}=5}%
    {\color{red} \repartitionsizes}
    {\color{black} \repartitionsizes}}}
\newcommand{\ShowMoveBoundaries}
{\moveboundaries}

\newcommand{\ShowBeforeUpdate}{\ifthenelse{\value{WSStep}<6}%
   {{\color{white} \beforeupdate}}
   {\ifthenelse{\value{WSStep}=6}%
    {\color{red} \beforeupdate}
    {\color{black} \beforeupdate}}}
\newcommand{\ShowAfterUpdate}{\ifthenelse{\value{WSStep}<7}%
   {{\color{white} \afterupdate}}
   {\ifthenelse{\value{WSStep}=7}%
    {\color{red} \afterupdate}
    {\color{black} \afterupdate}}}
\newcommand{\ShowUpdate}
{\update}

\newcommand{\FlaWorksheet}{
\begin{tabular}{| c | p{0.9\textwidth} |}\hline
Step & $\mbox{\color{blue}Algorithm:~}\routinename$
\\ \hline
\rowcolor{lightgray!25}   1a &%
$ \left\{
 \ShowPrecondition
\right\}
$%
\\ \hline
4 & \vspace{-0.15in}
\ShowPartitionings~ \\
&
\mbox{\color{blue} ~~~where~} \ShowPartitionSizes
\\ \hline
\rowcolor{lightgray!25}   
2 & 
$ \left\{ 
\ShowInvariant 
\right\} $ 
\\ \hline  
3 & $\mbox{\color{blue}while~} \ShowGuard \mbox{~\color{blue} do}$
\\ \hline 
\rowcolor{lightgray!25} 
2,3 &  
$ 
\left\{
\ShowInvariant \wedge \ShowGuardTwo
\right\}
$ 
\\ \hline 
5a & \vspace{-0.15in}
~~~~  \ShowRepartitionings
\\ \hline 
\rowcolor{lightgray!25} 6 & 
$ \left\{ 
\mbox{
~~~~ \ShowBeforeUpdate 
}
\right\}
$
\\ \hline 
8 & \vspace{-0.15in}
~~~~ \ShowUpdate
\\ \hline 
\rowcolor{lightgray!25} 7 & 
$ \left\{  
~~~~ \mbox{\ShowAfterUpdate }
\right\}
$
\\ \hline  
5b & \vspace{-0.15in} ~~~~ \mbox{
\ShowMoveBoundaries~
}
\\ \hline
\rowcolor{lightgray!25} 2 & 
$ \left\{ 
\mbox{
~~~~ $ \ShowInvariant  $ }
\right\}
$
\\ \hline
 
 &$\mbox{\color{blue} endwhile} $
\\ \hline 
\rowcolor{lightgray!25} 2,3 & 
$ \left\{ 
\mbox{
$ \ShowInvariant \wedge \neg( \ShowGuardTwo )$ 
}
\right\}
$
\\ \hline
\rowcolor{lightgray!25} 1b & 
$ \left\{ 
 \ShowPostcondition 
\right\}
$
\\ \hline
\end{tabular}
}

\newcommand{\FlaWorksheetWith}{
\begin{tabular}{| c | p{0.9\textwidth} |}\hline
Step & $\mbox{\color{blue}Algorithm:~}\routinename$
\\ \hline
\rowcolor{lightgray!25}   1a &%
$ \left\{ ~ P_{\rm pre}: \ShowPrecondition
~\right\}
$%
\\ \hline
4 & \vspace{-0.15in}
\ShowPartitionings~ 
\mbox{\color{blue} ~~~where~} \ShowPartitionSizes
\\ \hline
\rowcolor{lightgray!25}   
2 & 
$ \left\{ ~P_{\rm inv}: 
\ShowInvariant ~
\right\} $ 
\\ \hline  
3 &$\mbox{\color{blue}while~} \ShowGuard \mbox{~\color{blue} do}$
\\ \hline 
\rowcolor{lightgray!25} 
2,3 &  
$ 
~~~ \left\{ ~ P_{\rm inv} \wedge G:
 \ShowInvariant 
\wedge \ShowGuardTwo ~
\right\}
$ 
\\ \hline 
5a & \vspace{-0.15in}
~~~~  \ShowRepartitionings
\\ \hline 
\rowcolor{lightgray!25} 6 & 
$ ~~~ \left\{ ~ P_{\rm before}: 
\mbox{
~~~~ \ShowBeforeUpdate 
} ~
\right\}
$
\\ \hline 
8 & \vspace{-0.15in}
~~~~ \ShowUpdate
\\ \hline 
\rowcolor{lightgray!25} 7 & 
$ ~~~ \left\{  P_{\rm after}:  ~
 \mbox{\ShowAfterUpdate }
\right\}
$
\\ \hline  
5b & \vspace{-0.15in} ~~~~ \mbox{
\ShowMoveBoundaries~
}
\\ \hline
\rowcolor{lightgray!25} 2 & 
$ ~~~ \left\{ ~ P_{\rm inv}: 
\mbox{
 $ \ShowInvariant  $ }
\right\}
$
\\ \hline
 
 &$\mbox{\color{blue} endwhile} $
\\ \hline 
\rowcolor{lightgray!25} 2,3 & 
$  \left\{ ~ P_{\rm inv} \wedge \neg G:
\mbox{
$ \ShowInvariant \wedge \neg( \ShowGuardTwo )$ 
}
\right\}
$
\\ \hline
\rowcolor{lightgray!25} 1b & 
$ \left\{ ~ P_{\rm post}: 
 \ShowPostcondition 
~ \right\}
$
\\ \hline
\end{tabular}
}

\newcommand{\FlaWorksheetNine}{
\begin{tabular}{| c | p{0.9\textwidth} |}\hline
{\color{white}Step} & $\mbox{\color{blue}Algorithm:~}\routinename$
\\ \hline
 &%
$ \phantom{\left\{ 
\begin{minipage}{0.88\textwidth} 
$\ShowPrecondition$  
\end{minipage}
\right\}}
$%
\\ \hline
\rowcolor{lightgray!25}   
& %
\begin{minipage}{0.88\textwidth}%
\vspace{0.05in}
\ShowPartitionings~ \\
\mbox{\color{blue} ~~~where~} \ShowPartitionSizes
\end{minipage}
\\ \hline
& 
$ \phantom{\left\{ 
\begin{minipage}{0.88\textwidth} 
$\ShowInvariant $
\end{minipage}
\right\}} $ 
\\ \hline
\rowcolor{lightgray!25}   
&$\mbox{\color{blue}while~} \ShowGuard \mbox{~\color{blue} do}$
\\ \hline 
 &  
$
\phantom{\left\{
\begin{minipage}[t]{0.88\textwidth}%
~~~~$
\ShowInvariant 
\wedge \ShowGuardTwo$
\end{minipage}
\right\}}
$ 
\\ \hline
\rowcolor{lightgray!25}   
 & ~~~~ \begin{minipage}{0.85\textwidth}%
\vspace{0.05in}
\ifthenelse{\equal{\blocksize}{1}}{}%
{%
\ifthenelse{ \equal{\blocksize}{blank} }{}%
{{\bf Determine block size $ \blocksize $}\\}%
}
\ShowRepartitionings~ \\
$\mbox{\color{blue} ~~~where~}$ \ShowRepartitionSizes
\end{minipage}
\\ \hline
& 
$ \phantom{\left\{ 
\begin{minipage}{0.88\textwidth} 
~~~~ \ShowBeforeUpdate 
\end{minipage}
\right\}}
$
\\ \hline
\rowcolor{lightgray!25}  
 & ~~~~  \ShowUpdate 
\\ \hline 
& 
$ \phantom{\left\{ 
\begin{minipage}{0.88\textwidth} 
~~~~ \ShowAfterUpdate 
\end{minipage}
\right\}}
$
\\ \hline
\rowcolor{lightgray!25}   
 & ~~~~ \begin{minipage}{0.85\textwidth}%
\vspace{0.05in}
\ShowMoveBoundaries~
\end{minipage}
\\ \hline
& 
$ \phantom{\left\{ 
\begin{minipage}{0.88\textwidth} 
~~~~ $ \ShowInvariant  $ 
\end{minipage}
\right\}}
$
\\ \hline
\rowcolor{lightgray!25}  
 &$\mbox{\color{blue} endwhile} $
\\ \hline 
& 
$ \phantom{\left\{ 
\begin{minipage}{0.88\textwidth} 
$ \ShowInvariant \wedge \neg( \ShowGuardTwo )$ 
\end{minipage}
\right\}}
$
\\ \hline
& 
$ \phantom{\left\{ 
\begin{minipage}{0.88\textwidth} 
$ \ShowPostcondition $ 
\end{minipage}
\right\}}
$
\\ \hline
\end{tabular}
}

\newcommand{\FlaCostWorksheet}{
\begin{tabular}{| c | p{0.45\textwidth}
p{0.45\textwidth}|}\hline
Step & $\mbox{\color{blue}Algorithm:~}\routinename $ &
\\ \hline
1a &%
$ \left\{ 
\begin{minipage}{0.44\textwidth} 
$\ShowPrecondition$  
\end{minipage}
\right\}
$
&
\\ \hline
\rowcolor{lightgray!25}   
4 & %
\begin{minipage}{0.88\textwidth}%
\vspace{0.05in}
\ShowPartitionings~ \\
\mbox{\color{blue} ~~~where~} \ShowPartitionSizes
\end{minipage}
& 
\begin{minipage}{0.44\textwidth}
\hfill \CostInit
\end{minipage}
\\ \hline
2 & 
$ \left\{ 
\begin{minipage}{0.44\textwidth} 
$\ShowInvariant $
\end{minipage}
\right\} $ 
&
\begin{minipage}{0.44\textwidth}
$ \{ $ \hfill \CostInvariant
$ \} $ 
\end{minipage}
\\ \hline
\rowcolor{lightgray!25}   
3 &$\mbox{\color{blue}while~} \ShowGuard \mbox{~\color{blue} do}$
&
\\ \hline 
2,3 &  
$
\left\{
\begin{minipage}[t]{0.41\textwidth}%
$
~~~~ \ShowInvariant 
\wedge \ShowGuardTwo$
\end{minipage}
\right\}
$ 
&
\begin{minipage}{0.44\textwidth}
$ \{ $ \hfill \CostInvariant
$ \} $
\end{minipage}
\\ \hline
\rowcolor{lightgray!25}   
5a & ~~~~ \begin{minipage}{0.41\textwidth}%
\vspace{0.05in}
\ifthenelse{\equal{\blocksize}{1}}{}%
{%
\ifthenelse{ \equal{\blocksize}{blank} }{}%
{{\bf Determine block size $ \blocksize $}\\}%
}
\ShowRepartitionings~ 
\NoShow{\\
$\mbox{\color{blue} ~~~where~}$ \ShowRepartitionSizes}
\end{minipage}
&
\\ \hline
6 & 
$ \left\{ 
\begin{minipage}{0.41\textwidth} 
~~~~ \ShowBeforeUpdate 
\end{minipage}
\right\}
$
&
$ \{ $ \hfill 
\CostBefore
$ \} $
\\ \hline
\rowcolor{lightgray!25}  
8 & 
\begin{minipage}{0.41\textwidth} 
~~~~  \ShowUpdate 
\end{minipage}
&
\hfill 
$
C := C + 2 
$
\\ \hline 
7 & 
$ \left\{ 
\begin{minipage}{0.41\textwidth} 
~~~~ \ShowAfterUpdate 
\end{minipage}
\right\}
$
&
$ \{ $ \hfill 
\CostAfter
$ \} $
\\ \hline
\rowcolor{lightgray!25}   
5b & ~~~~ \begin{minipage}{0.41\textwidth}%
\vspace{0.05in}
\ShowMoveBoundaries~
\end{minipage}
&
\\ \hline
2 & 
$ \left\{ 
\begin{minipage}{0.44\textwidth} 
~~~~ $ \ShowInvariant  $ 
\end{minipage}
\right\}
$
&
\begin{minipage}{0.44\textwidth}
$ \{ $ \hfill \CostInvariant
$ \} $
\end{minipage}
\\ \hline
\rowcolor{lightgray!25}  
 &$\mbox{\color{blue} endwhile} $
 &
\\ \hline 
2,3 & 
$ \left\{ 
\begin{minipage}{0.44\textwidth} 
$ \ShowInvariant \wedge \neg( \ShowGuardTwo )$ 
\end{minipage}
\right\}
$
&
\begin{minipage}{0.44\textwidth}
$ \{ $ \hfill \CostInvariant
$ \} $
\end{minipage}
\\ \hline
1b & 
$ \left\{ 
\begin{minipage}{0.44\textwidth} 
$ \ShowPostcondition $ 
\end{minipage}
\right\}
$
&
\begin{minipage}{0.44\textwidth}
$ \{ $ \hfill \CostPostCond
$ \} $
\end{minipage}
\\ \hline
\end{tabular}
}

\usepackage{microtype}

\newtheorem{theorem}
{Theorem}[section]

\newtheorem{corollary}[theorem]
{Corollary}
\newtheorem{definition}[theorem]
{Definition}
\definecolor{FutureColor}{gray}{0.6}

\newcommand{\future}[1]{{\widetilde #1}}

\title{Enabling Pivoting in the Formal Derivation of LU factorization}

\author{Robert A. van de Geijn \\
Margaret E. Myers \\
Devangi N. Parikh \\
The University of Texas at Austin \\
\href{mailto:rvdg@cs.utexas.edu}{rvdg@cs.utexas.edu},
\href{mailto:myers@cs.utexas.edu}{myers@cs.utexas.edu},
\href{mailto:dnp@cs.utexas.edu}{dnp@cs.utexas.edu}
\\[0.2in]
Tze Meng Low \\
Carnegie Mellon University \\
\href{mailto:lowt@andrews.cmu.edu}{lowt@andrews.cmu.edu}
\\[0.2in]
Devin A. Matthews \\
Southern Methodist University \\
 \href{mailto:damatthews@smu.edu}{damatthews@smu.edu}
}

\date{\today}

\allowdisplaybreaks  

\begin{document}

\maketitle

\begin{abstract}
The FLAME methodology for deriving linear algebra algorithms from  specification, first introduced around 2000,  has been successfully applied to a broad cross section of operations. An open question has been whether it can yield algorithms for the best-known operation in linear algebra, LU factorization with partial pivoting (Gaussian elimination with row swapping).
This paper shows that it can and provides general techniques for pivoted factorizations.

\end{abstract}

\section{Introduction}

The Formal Linear Algebra Methods Environment (FLAME) methodology for deriving linear algebra algorithms was first presented in 2000 at the
IFIP TC2/WG2.5 Working Conference on the Architecture of Scientific Software and published in related early works in 2001~\cite{Gunnels:PhD,FLAME,FLAME_WoCo}.  
The approach makes algorithm discovery  a goal-oriented (correct-by-design) endeavor that starts with the specification of the input (as a precondition) and output (as a postcondition.) From  these, it
derives 
algorithms
hand in hand with their proofs of correctness.
 While the resulting algorithms are loop based, they can incorporate recursion since the target operation appears as a suboperation, thus creating a rich space of algorithms from which one with high performance (or other desirable attribute) can be chosen.

Since its inception, the approach was reformulated into a worksheet~\cite{Recipe}, applied to a broad class of operations~\cite{Bientinesi2013,Bientinesi:2008:FAR,Quintana-Orti:2003:FDA,TSoPMC},
\NoShow{implemented as a mechanical system~\cite{Paolo:PhD,Cl1ck},} employed to develop parts of the widely-used dense linear algebra library (libflame)~\cite{libflame_ref,CiSE09}, and extended to systematically derive backward error analyses~\cite{Bientinesi:2011:GMS:2078718.2078728}.
Its scope has expanded to include  Krylov subspace methods (iterative methods for solving linear systems)~\cite{Eijkhout20101805} and 
graph operations represented with matrices~\cite{9835383,TC_Correctness}.
In~\cite{phd:low}, an argument is made that  the underlying principles apply to primitive recursive functions.
A comprehensive discussion of this research and its impact is given in a chapter~\cite{10.1145/3544585.3544597}
in a book honoring Edsger W. Dijkstra~\cite{10.1145/3544585}, whose vision greatly influenced the work.

Dense LU factorization with partial pivoting ({\sc LUpiv}) is possibly the most studied and among the most important operations in linear algebra. A 25 year old open question of whether FLAME applies to it (in contrast to un-pivoted LU), first posed by Jim Demmel, is finally answered here.
More broadly, many operations require pivoting for numerical stability, such as Cholesky factorization with symmetric pivoting, QR factorization with column pivoting, and tridiagonal decomposition via Gaussian transforms (with pivoting) of symmetric and  skew-symmetric matrices~\cite{LTLt}.  This paper provides general strategies that can unlock the FLAME methodology for such operations.

When the FLAME methodology was first proposed, it was already noted that it is very systematic and that hence correct-by-design derivation of linear algebra algorithms could be made mechanical (automatic). A prototype system was developed~\cite{Paolo:PhD}, which was eventually refined into Cl1ck~\cite{Cl1ck}.
Recently, FLAME-like notation was also used to prove theorems about the LU (without pivoting) and Cholesky factorizations with the ACL2 automatic theorem prover~\cite{10918400,kwan_et_al:LIPIcs.ITP.2024.25}.
We have written this paper in a way that exposes  knowledge in support of automation.  

Together, this work moves the discovery of algorithms in dense linear algebra further from being an art performed by a human expert to a science that lays the foundation for knowledge synthesis and automation.

\NoShow{
In Section~\ref{sec:LUnopiv}, we use LU factorization without pivoting (LUnopiv) to review the methodology.  In Section~\ref{sec:LUpiv}, we show how pivoting can be accommodated.  In Section~\ref{sec:conclusion}, we discuss how the insights gained from this study reveal LU with pivoting to be an edge case for the FLAME methodology.
}

\section{Deriving algorithms for LU factorization without pivoting.}
\label{sec:LUnopiv}

We start with the LU factorization without pivoting, {\sc LUnopiv}:
Given an $ n \times n $ matrix $ A $, compute unit lower triangular matrix $ L $ and upper triangular matrix $ U $ (with nonzero diagonal elements) such that 
$ A = L U $. 
Square matrices are used here for simplicity, although the approach can be straightforwardly applied to rectangular matrices as well.
This factorization exists if and only if all principal major submatrices of $ A $ are nonsingular.  
For more detailed derivations, see~\cite{LAFF-On-Correctness,TSoPMC}.  What is new here is the alternative way in which loop invariants are presented, in Section~\ref{sec:altinv}.

We embrace ``Householder notation,'' where matrices are denoted with upper case Roman letters, column vectors with lower case Roman letters, and scalars with lower case Greek letters. 

\subsection{The FLAME notation}

\begin{figure}[tb!]
\input LU_unb_all

\begin{center}
  \FlaAlgorithm  
\end{center}
\caption{Five  unblocked algorithms for {\sc LUnopiv}.  Matrices $ L_{00} $ and $ U_{00} $ are the unit lower triangular and upper triangular matrices computed in prior iterations and stored in $ A_{00} $.  $ a_{01} := L_{00}^{-1} a_{01} $ and $ a_{10}^T := a_{10}^T U_{00}^{-1} $ are implemented as triangular solves.}
\label{fig:LU_unb_all}
\end{figure}

The hiding of the index details using the FLAME notation is key to the FLAME approach~\cite{FLAME}.
In Figure~\ref{fig:LU_unb_all}, we present all classical unblocked algorithms for {\sc LUnopiv} using this notation.

\subsection{Proving correctness}
\label{sec:proving_correctness}

\begin{figure}[tbp]

\input LU_unb_var3_ws

\begin{center}
\setlength{\arraycolsep}{2pt}
\FlaWorksheetWith
\end{center}

\caption{Annotated left-looking algorithm.  When this ``worksheet'' is filled out in the order indicated by the column marked ``Step,'' it enables the hand-in-hand derivation the algorithm and its proof of correctness. }
\label{fig:left-ws}
\end{figure}

In Figure~\ref{fig:left-ws}, the FLAME worksheet~\cite{Recipe} for the left-looking algorithm (Variant 3) is given,  annotated with predicates (in the gray boxes) that describe the state of the variables at the indicated points. 
The worksheet
provides a framework for proving a loop correct 
using the {While Theorem}~\cite{Dijkstra,Gries,10.1145/363235.363259,LAFF-On-Correctness}, which itself relies on the Principle of Mathematic Induction (PMI).
The commands in the algorithm have the property that they maintain the truth of  these assertions, thus verifying the correctness of the algorithm, in the spirit of Hoare Logic~\cite{10.1145/363235.363259,Gries}.
Here, $ \widehat A $ equals the original content of $ A $.  The notation $ A = L \backslash U $ captures that upon completion $ A $ has been overwritten by $ L $ below the diagonal (with the diagonal implicitly stored) and $ U $ on and above the diagonal.
For  brevity in the predicates, the existence and triangular nature of $ L $ and $ U $ are implicitly assumed.

The precondition (Step~1a) and postcondition (Step~1b) specify input and output.
Critical to the proof of correctness is the {\em loop invariant},
\[
\FlaTwoByTwo{A_{TL}}{A_{TR}}
               {A_{BL}}{A_{BR}} =
  \FlaTwoByTwo{L \backslash U_{TL}}{\widehat{A}_{TR}}
               {L_{BL}}{\widehat{A}_{BR}}
               \wedge
               \FlaTwoByOne{ \widehat A_{TL}}
               {\widehat A_{BL} }
               =
               \FlaTwoByOne{ L_{TL} U_{TL}}
               { L_{BL} U_{TL} },
\]
which captures the state of the matrix $ A $ before and after each iteration (at the four places marked Step~2.)
The loop guard in Step~3 has the property that when it becomes {\em false} (meaning control exits the loop,) the loop invariant implies the postcondition. The initialization in Step 4 leaves the variables in the state described by the loop invariant before the start of the loop, meaning it is true at the top of the loop body in the first iteration and providing  
the base case for a proof by induction.
The assertions in the loop body can be used to reason that if the loop invariant holds at the top of the loop body, then it again holds at the bottom of the loop body,
providing the inductive step in a proof by induction.
By the PMI, we conclude that the loop invariant holds at the top and bottom of the loop body every time the loop  executes.
Thus, when control leaves the loop (the loop guard  becomes {\em false}), the loop invariant still holds after the loop.  The loop invariant being {\em true} and the loop guard being {\em false} together imply the postcondition (that the desired result has been computed.)
Hence, this loop correctly computes the LU factorization of the original matrix $ A $ {\em if} the loop finishes.  The loop finishes because in every iteration the computation progresses forward through the finite-sized matrix $ A$ and it is assumed $ A $ has an LU factorization.

\subsection{Deriving families of correct algorithms}
\label{sec:LUnopiv:derivation}

\begin{figure}[tb!]
\setlength{\arraycolsep}{2pt}

\input blanksheet

\FlaWorksheet
\caption{Blank worksheet}
\label{fig:blank}
\end{figure}

While proving a loop correct is useful, Dijkstra~\cite{EWD:EWD340} advocated developing the loop hand in hand with its proof of correctness from the precondition  and the postcondition.  
This means we  derive a family of algorithms by starting with the empty worksheet in Fig.~\ref{fig:blank}, filling it step-by-step until the algorithm and its proof of correctness result.

The process starts by stating the precondition, $ A = \widehat A $, and postcondition, $ A = L \backslash U  \wedge \widehat A = L U  $.
These predicates are added to the worksheet as Steps~1a and~1b, as illustrated in Figure~\ref{fig:left-ws}.
Loop invariants can be derived from the precondition and postcondition, starting by partitioning the matrices $A$, $L$, and $U$. We observe that $ L $ and $ U $ should be partitioned into quadrants to expose the blocks of zeroes as well as the triangular nature of the top-left and bottom-right submatrices.  This means $ A $ itself also needs to be partitioned into quadrants, since $ L $ and $ U $ overwrite $ A $ and since $ \widehat A = L U $ forces a conformal partitioning on $ \widehat A $.  These partitioned matrices are substituted into the postcondition, yielding
\begin{equation}
\label{eqn:substitute}
\FlaTwoByTwo
{ A_{TL}} {A_{TR}}
{ A_{BL}} {A_{BR}}
=
\FlaTwoByTwo
{ L \backslash U_{TL}} {U_{TR}}
{ L_{BL}} {L \backslash U_{BR}}
\wedge
\FlaTwoByTwo
{ \widehat A_{TL}} {\widehat A_{TR}}
{ \widehat A_{BL}} {\widehat A_{BR}}
=
\FlaTwoByTwo
{ L_{TL}} {0}
{ L_{BL}} {L_{BR}}
\FlaTwoByTwo
{ U_{TL}} {U_{TR}}
{ 0 } {U_{BR}}.
\end{equation}
Multiplying the partitioned $ L $ and $ U $, and manipulating the result, gives us
\begin{equation}
\label{eqn:substitute2}
\FlaTwoByTwo
{ A_{TL}} {A_{TR}}
{ A_{BL}} {A_{BR}}
=
\FlaTwoByTwo
{ L \backslash U_{TL}} {U_{TR}}
{ L_{BL}} {L \backslash U_{BR}}
\wedge
\FlaTwoByTwo
{ \widehat A_{TL}} {\widehat A_{TR}}
{ \widehat A_{BL}} {\widehat A_{BR} - L_{BR} U_{BR}}
=
\FlaTwoByTwo
{ L_{TL} U_{TL} } { L_{TL} U_{TR}}
{L_{BL} U_{TL}} {L_{BL} U_{TR} },
\end{equation}
which is a recursive definition for {\sc LUnopiv}.
We call this the {\em Partitioned Math Expression} (PME) for {\sc LUnopiv}.

\begin{figure}[tb!]
\setlength{\arraycolsep}{2pt}

\begin{center}
\begin{tabular}{| c | l @{~} l |} \hline
Invariant & 
\multicolumn{2}{c|}{$ P_{\rm inv}$}  \\ \whline
&& \mypadding
1
&
\small
$\FlaTwoByTwo
{ A_{TL}} {A_{TR}}
{ A_{BL}} {A_{BR}}
=
\FlaTwoByTwo
{ L \backslash U_{TL}} {\widehat A_{TR}}
{ \widehat A_{BL}} {\widehat A_{BR}}
$
&
\small
$
\wedge
\FlaTwoByTwo
{ \widehat A_{TL}} {\phantom{\widehat A_{TR}}}
{ \phantom{\widehat A_{BL}}} {}
=
\FlaTwoByTwo
{ L_{TL} U_{TL} } { \phantom{L_{TL} U_{TR}}}
{\phantom{L_{BL} U_{TL}}} {}
$
\\ && \mypadding \hline
&& \mypadding
2
&
\small
$
\FlaTwoByTwo
{ A_{TL}} {A_{TR}}
{ A_{BL}} {A_{BR}}
=
\FlaTwoByTwo
{ L \backslash U_{TL}} {U_{TR}}
{ \widehat A_{BL}} {\widehat A_{BR}}
$
&
\small
$
\wedge
\FlaTwoByTwo
{ \widehat A_{TL}} {\widehat A_{TR}}
{ \phantom{\widehat A_{BL}}} {\phantom{\widehat A_{BR}}}
=
\FlaTwoByTwo
{ L_{TL} U_{TL} } { L_{TL} U_{TR}}
{\phantom{L_{BL} U_{TL}}} {}
$
\\ && \mypadding \hline
&& \mypadding
3
&
\small
$
\FlaTwoByTwo
{ A_{TL}} {A_{TR}}
{ A_{BL}} {A_{BR}}
=
\FlaTwoByTwo
{ L \backslash U_{TL}} { \widehat A_{TR}}
{ L_{BL}} { \widehat A_{BR} }
$
&
\small
$
\wedge
\FlaTwoByTwo
{ \widehat A_{TL}} {\phantom{\widehat A_{TR}}}
{ \widehat A_{BL}} {\phantom{\widehat A_{BR}}}
=
\FlaTwoByTwo
{ L_{TL} U_{TL} } { \phantom{L_{TL} U_{TR}}}
{L_{BL} U_{TL}} {}
$
\\ && \mypadding \hline
&& \mypadding
4
&
\small
$
\FlaTwoByTwo
{ A_{TL}} {A_{TR}}
{ A_{BL}} {A_{BR}}
=
\FlaTwoByTwo
{ L \backslash U_{TL}} {U_{TR}}
{ L_{BL}} {\widehat A_{BR}}
$
&
\small
$
\wedge
\FlaTwoByTwo
{ \widehat A_{TL}} {\widehat A_{TR}}
{ \widehat A_{BL}} {\phantom{\widehat A_{BR}}}
=
\FlaTwoByTwo
{ L_{TL} U_{TL} } { L_{TL} U_{TR}}
{L_{BL} U_{TL}} {}
$
\\ && \mypadding \hline
&& \mypadding
5
&
\small
$
\FlaTwoByTwo
{ A_{TL}} {A_{TR}}
{ A_{BL}} {A_{BR}}
=
\FlaTwoByTwo
{ L \backslash U_{TL}} {U_{TR}}
{ L_{BL}} {\widehat A_{BR} -  L_{BL} U_{TR}} 
$
&
\small
$
\wedge
\FlaTwoByTwo
{ \widehat A_{TL}} {\widehat A_{TR}}
{ \widehat A_{BL}} {}
=
\FlaTwoByTwo
{ L_{TL} U_{TL} } { L_{TL} U_{TR}}
{L_{BL} U_{TL}} {}
$
\mypadding\ &&

\\ \hline
\end{tabular}
\end{center}
    \caption{Five loop invariants for LU factorization (without pivoting).}
    \label{fig:LUnopiv_invs}
\end{figure}

Now we transform the recursive definition into loop-based algorithms  
by extracting loop invariants from the PME.  
We recognize that the PME captures {\em all} computation that must be performed to leave the variables in a state where the postcondition is {\em true}.  
At the top and bottom of a typical iteration,
variables contain a partial result.
  This guides the derivation of loop invariants from the PME: keep some unique set of subexpressions
  from the PME that capture this partial progress, constrained by the dependence between subexpressions.
Applying these principles yields the five loop invariants in Fig.~\ref{fig:LUnopiv_invs}.

In Fig.~\ref{fig:blank}, we now choose $ P_{\rm inv} $ (in each instance of Step~2) to equal Invariant~3, the loop invariant for the left-looking algorithm, so that upon completion the worksheet will mirror that in Fig.~\ref{fig:left-ws}.  
Next, in Step~3, a {\em loop guard}, $ G $, can be determined from the fact that $ P_{\rm inv} \wedge \neg G $ must imply the postcondition, yielding $ n( A_{TL} ) < n( A ) $, where the function $n(\cdot)$ counts the number of columns in a (sub-)matrix.
Similarly, the initialization in Step~4 follows from the fact that it must take the variables from the state described by the precondition to the state where the loop invariant holds: $ A_{TL} $, $ L_{TL} $, and $ U_{TL} $ must be $ 0 \times 0 $.
Next, in Step~5 it is determined how to expose the parts of the matrices that will be updated in the current iteration by further partitioning the matrix, which also determines how progress is made. 
In Step~6, the state of these exposed parts, before any updates are applied, is determined by substituting the exposed submatrices into the loop invariant and manipulating the result.
Similarly, in Step~7, the state that the exposed parts must acquire for the invariant to again hold at the bottom of the loop is determined. 
Finally, assignments are deduced that take the state of the variables from that described by $ P_{\rm before} $ to that described by $ P_{\rm after} $, in Step~8.
The completed worksheet in Figure~\ref{fig:left-ws} is the result.  
By removing the annotations that are part of the proof of correctness, Variant~3 (the left-looking algorithm) in Figure~\ref{fig:LU_unb_all} remains.
This process can be repeated to derive all five variants given in Figure~\ref{fig:LU_unb_all}.

Details on the FLAME methodology in general, and LU without pivoting in particular, can be found in~\cite{LAFF-On-Correctness,TSoPMC}.  There, it is also shown that all five algorithms perform the same number of floating point operations.  A discussion of how to derive blocked algorithms that cast most computation in terms of matrix-matrix multiplication and can achieve high performance can be found there as well.

\subsection{Alternative formulation}
\label{sec:altinv}

The described formal derivation process makes the discovery of algorithms a goal-oriented activity~\cite{Gries}.  
As we started  applying it to {\sc LUpiv}, it became clear that we needed to specify invariants differently.
This is because stating the loop invariant only in terms of data currently available does not fit with the fact that all pivoting information is not yet available for a typical loop iteration.
We now first describe an alternative in the setting of {\sc LUnopiv}.

We again use the derivation of the unblocked left-looking algorithm (Variant~3) as an example.
In~Sections~\ref{sec:proving_correctness} and~\ref{sec:LUnopiv:derivation}, the precondition is given as $ A = \widehat A $. What this does not capture is that the algorithm will only  successfully complete the loop if the LU factorization exists.
This more complete precondition
is captured by 
$
A = \widehat A \wedge ( \exists L, U ~ \vert ~ \widehat A = L U )
$
or, more concisely,
\[
A = \widehat A 
\wedge 
\widehat A = 
\future{ L}\future{  U},
\]
where $ \widehat A = \future{ L}\future{  U}
$ indicates that ``there exist $ L $ and $ U $ such that $ \widehat A = LU $, but those $ L $ and $ U $ have not yet been computed.''  
The triangular nature of $ L $ and $ U $ continues to be implicitly assumed.

The postcondition continues to be given by 
$
A = L \backslash U 
\wedge \widehat A = L U $.
Substituting the partioned matrices into this postcondition again yields the PME in~\ref{eqn:substitute}.
We now manipulate this to yield an expression that is an alternative to~(\ref{eqn:substitute2}):
\begin{equation}
\nonumber
\FlaTwoByTwo
{ A_{TL}} {A_{TR}}
{ A_{BL}} {A_{BR}}
=
\FlaTwoByTwo
{ L \backslash U_{TL}} {U_{TR}}
{ L_{BL}} {L \backslash U_{BR}}
\wedge
\begin{array}[t]{c}
\underbrace{
\FlaTwoByTwo
{ \widehat A_{TL}} {\widehat A_{TR}}
{ \widehat A_{BL}} {\widehat A_{BR} }
=
\FlaTwoByTwo
{ L_{TL} U_{TL} } { L_{TL} U_{TR}}
{L_{BL} U_{TL}} {L_{BL} U_{TR} + L_{BR} U_{BR} }
}
\\
\mbox{constraint}
\end{array}
.
\end{equation}
Invariant~3  for the left-looking algorithm can now be presented instead as
\begin{eqnarray*}
\FlaTwoByTwo
{ A_{TL}} {A_{TR}}
{ A_{BL}} {A_{BR}}
&=&
\FlaTwoByTwo
{ L \backslash U_{TL}}
{L_{TL} \future{  U_{TR}}}
{ L_{BL}} {L_{BL} 
\future{ U_{TR}} + \future{ L_{BR} }\future{ U_{BR}}} \\
&\wedge&
\begin{array}[t]{c}
\underbrace{
\FlaTwoByTwo
{ \widehat A_{TL}} {\widehat A_{TR}}
{ \widehat A_{BL}} {\widehat A_{BR} }
=
\FlaTwoByTwo
{ L_{TL} U_{TL} } { L_{TL} \future{  U_{TR}}}
{L_{BL} U_{TL}} {L_{BL} \future{ U_{TR}} + \future{ L_{BR}}\future{  U_{BR} }}}
\\
\mbox{constraint}
\end{array}
,
\end{eqnarray*}
where some parts of the expressions have not yet been computed.

As before, Steps 3--5 can be derived and are identical to those in Figure~\ref{fig:left-ws}.  The state in Step~6 is now 
{\setlength{\arraycolsep}{2pt}
\[
\left( \begin{array}{c I c }
A_{00} & 
\left( \begin{array}{c | c }
a_{01} & A_{02} 
\end{array} \right)
\\ \mypadding[-2pt] \whline
\mypadding[-2pt]
\left( \begin{array}{c  }
a_{10}^T \\ \hline
A_{20} 
\end{array} \right)
&
\left( \begin{array}{c | c }
\alpha_{11} & a_{12}^T  \\ \hline
a_{21} & A_{22} 
\end{array} \right)
\end{array} \right) =
\left( \begin{array}{c I c }
L \backslash U_{00} & 
L_{00} 
\left( \begin{array}{c | c }
\future{ u_{01}} & \future{  U_{02} }
\end{array} \right)
\\ \mypadding[-2pt] \whline
\mypadding[-2pt]
\left( \begin{array}{c  }
l_{10}^T \\ \hline
L_{20} 
\end{array} \right)
&
\left( \begin{array}{c  }
l_{10}^T \\ \hline
L_{20} 
\end{array} \right)
\left( \begin{array}{c | c }
\future{ u_{01}} & \future{ U_{02} }
\end{array} \right)
+
\left( \begin{array}{c | c }
1 & 0 \\ \hline
\future{ l_{21}} & \future{ L_{22} }
\end{array} \right)
\left( \begin{array}{c | c }
\future{ \upsilon_{11}} & \future{ u_{12}^T } \\ \hline
0 & \future{ U_{22} }
\end{array} \right)
\end{array} \right)
\wedge \cdots 
\]%
}
or, equivalently,
\[
\left( \begin{array}{c I c | c }
A_{00} & 
a_{01} & A_{02} 
\\ \whline
a_{10}^T & \alpha_{11} & a_{12}^T  \\ \hline
A_{20} &
a_{21} & A_{22} 
\end{array} \right) =
\left( \begin{array}{c I c | c }
L \backslash U_{00} & 
L_{00} \future{ u_{01}} & L_{00} \future{  U_{02}} 
\\ \whline
l_{10}^T & l_{10}^T \future{  u_{01}^T} + \future{   \upsilon_{11}} & l_{10}^T \future{ U_{02}} + \future{ u_{12}^T }\future{ a_{12}^T}  \\ \hline
L_{20} &
L_{20} \future{ u_{01}} + \future{ \upsilon_{11}}\future{  l_{21}} & L_{20} \future{ U_{02}} +  \future{ l_{21} }\future{ u_{12}^T} + \future{ L_{22}}\future{  U_{22}}
\end{array} \right)
\wedge
\mbox{constraint}.
\]
In Step~7 the following must hold:
\[
\left( \begin{array}{c I c }
\left( \begin{array}{c | c }
A_{00} & a_{01} \\ \hline
a_{10}^T & \alpha_{11}
\end{array} \right)
&
\left( \begin{array}{c }
A_{02} \\ \hline
a_{12}^T
\end{array} \right)
\\ \mypadding[-2pt]\whline
\mypadding[-2pt]
\left( \begin{array}{c | c  }
A_{20} & a_{21}
\end{array} \right)
&
A_{22}
\end{array} \right) 
=
\left( \begin{array}{c I c }
\left( \begin{array}{c | c }
L \backslash U_{00} &  u_{01} \\ \hline
l_{10}^T & \upsilon_{11}
\end{array} \right)
&
\left( \begin{array}{c | c }
L_{00} & 0 \\ \hline
l_{10}^T & 1
\end{array} \right) 
\left( \begin{array}{c }
\future{ U_{02}} \\ \hline
\future{ u_{12}^T}
\end{array} \right)
\\ \mypadding[-2pt]\whline
\mypadding[-2pt]
\left( \begin{array}{c | c  }
L_{20} & l_{21}
\end{array} \right)
&
\left( \begin{array}{c | c  }
L_{20} & l_{21}
\end{array} \right)
\left( \begin{array}{c }
\future{ U_{02}} \\ \hline
\future{ u_{12}^T}
\end{array} \right)
+
\future{ L_{22} }\future{ U_{22}}
\end{array} \right)
\wedge
\mbox{constraint}
\]
or, equivalently,
\[
\left( \begin{array}{c | c I c }
A_{00} & 
a_{01} & A_{02} 
\\ \hline
a_{10}^T & \alpha_{11} & a_{12}^T  \\ \whline
A_{20} &
a_{21} & A_{22} 
\end{array} \right) =
\left( \begin{array}{c | c I c }
L \backslash U_{00} & 
u_{01} & L_{00} \future{  U_{02} }
\\ \hline
l_{10}^T &  \upsilon_{11} & l_{10}^T \future{  U_{02}} + \future{ u_{12}^T }\future{ U_{02}^T}  \\ \whline
L_{20} & l_{21} & L_{20} \future{ U_{02}} +  l_{21} \future{ u_{12}^T} + \future{ L_{22}}\future{  U_{22}}
\end{array} \right)
\wedge
\mbox{constraint}.
\]
Comparing the states in Steps~6 and~7 prescribes the updates
\[
\begin{array}{l}
a_{01} := L_{00}^{-1} a_{01} ~~~~ \mbox{(triangular solve)} \\
\alpha_{11} := \alpha_{11} - a_{10}^T a_{01} \\
a_{21} := a_{21} - A_{20} a_{01} \\
a_{21} := a_{21} / \alpha_{11}.
\end{array}
\]

In this alternative way of presenting the PME and resulting loop invariant, the goal (the final result) shows up in the description of what is currently contained in $ A $.
This  will make it possible to derive algorithms that include pivoting.  

In Figure~\ref{fig:LUnopiv_invs2} we compare and contrast how the contents of $ A $ are captured for each of the five loop invariants with the old and new formulations.

\begin{figure}[tb!]
\setlength{\arraycolsep}{2pt}

\begin{center}
\begin{tabular}{| c | l | l |} \hline
& \multicolumn{2}{c|}{Contents of $ A $} \\ \cline{2-3}
Invariant & 
Old formulation & New formulation  \\ \whline
&& \mypadding
1
&
\small
$\FlaTwoByTwo
{ A_{TL}} {A_{TR}}
{ A_{BL}} {A_{BR}}
=
\FlaTwoByTwo
{ L \backslash U_{TL}} {\widehat A_{TR}}
{ \widehat A_{BL}} {\widehat A_{BR}}
$
&
\small
$\FlaTwoByTwo
{ A_{TL}} {A_{TR}}
{ A_{BL}} {A_{BR}}
=
\FlaTwoByTwo
{ L \backslash U_{TL}} { L_{TL}\future{ U_{TR}}}
{ \future{ L_{BL}}  U_{TL} } {\future{L_{BL}}\future{ U_{TR}} + \future{ L_{BR}}\future{ U_{BR}} }
$
\\ && \mypadding \hline
&& \mypadding
2
&
\small
$
\FlaTwoByTwo
{ A_{TL}} {A_{TR}}
{ A_{BL}} {A_{BR}}
=
\FlaTwoByTwo
{ L \backslash U_{TL}} {U_{TR}}
{ \widehat A_{BL}} {\widehat A_{BR}}
$
&
\small
$\FlaTwoByTwo
{ A_{TL}} {A_{TR}}
{ A_{BL}} {A_{BR}}
=
\FlaTwoByTwo
{ L \backslash U_{TL}} { U_{TR}}
{ \future{ L_{BL}}  U_{TL} } {\future{ L_{BL}}  U_{TR} + \future{ L_{BR}}\future{  U_{BR}} }
$
\\ && \mypadding \hline
&& \mypadding
3
&
\small
$
\FlaTwoByTwo
{ A_{TL}} {A_{TR}}
{ A_{BL}} {A_{BR}}
=
\FlaTwoByTwo
{ L \backslash U_{TL}} { \widehat A_{TR}}
{ L_{BL}} { \widehat A_{BR} }
$
&
\small
$\FlaTwoByTwo
{ A_{TL}} {A_{TR}}
{ A_{BL}} {A_{BR}}
=
\FlaTwoByTwo
{ L \backslash U_{TL}} {L_{TL} \future{ U_{TR}}}
{ L_{BL} } {L_{BL} \future{ U_{TR}} + \future{ L_{BR} \future{ U_{BR}}} }
$
\\ && \mypadding \hline
&& \mypadding
4
&
\small
$
\FlaTwoByTwo
{ A_{TL}} {A_{TR}}
{ A_{BL}} {A_{BR}}
=
\FlaTwoByTwo
{ L \backslash U_{TL}} {U_{TR}}
{ L_{BL}} {\widehat A_{BR}}
$
&
\small
$\FlaTwoByTwo
{ A_{TL}} {A_{TR}}
{ A_{BL}} {A_{BR}}
=
\FlaTwoByTwo
{ L \backslash U_{TL}} {  U_{TR}}
{ L_{BL}  } {L_{BL} U_{TR} + \future{ L_{BR} }\future{ U_{BR}} }
$
\\ && \mypadding \hline
&& \mypadding
5
&
\small
$
\FlaTwoByTwo
{ A_{TL}} {A_{TR}}
{ A_{BL}} {A_{BR}}
=
\FlaTwoByTwo
{ L \backslash U_{TL}} {U_{TR}}
{ L_{BL}} {\widehat A_{BR} -  L_{BL} U_{TR}} 
$
&
\small
$\FlaTwoByTwo
{ A_{TL}} {A_{TR}}
{ A_{BL}} {A_{BR}}
=
\FlaTwoByTwo
{ L \backslash U_{TL}} {U_{TR}}
{ L_{BL}} {\future{ L_{BR} }\future{ U_{BR} }}
$
\mypadding\ &&
\\ \hline
\end{tabular}
\end{center}
    \caption{Five loop invariants for LU factorization (without pivoting).  
    }
    \label{fig:LUnopiv_invs2}
\end{figure}

\section{Deriving algorithms for LU factorization with pivoting}
\label{sec:LUpiv}

A classic result in numerical analysis is that, due to roundoff error, {\sc LUnopiv} computes  approximate factors that equal the exact decomposition of a perturbed matrix. 
More formally,
if $ A $ is $ n \times n $ and in exact arithmetic $  A = L U $, then the computed factors $ \check L  $ and $ \check U $ 
satisfy 
$ A + E = \check L \check U $,
where $ \vert E \vert \leq c_0 \epsilon_{\rm mach} \vert \check L \vert \vert \check U \vert +
c_1 \epsilon_{\rm mach} \vert A \vert
$ with constants $ c_0 $ and $ c_1 $ that depend on $ n $ and the algorithmic variant that is employed~\cite{Bientinesi:2011:GMS:2078718.2078728,GVL4,Higham:2002:ASN}.
Here $ \vert X \vert $ equals the matrix of element-wise absolute values and $ X \leq Y $ is the conjunction of element-wise comparisons.
The key insight is that if elements of $ \check L $ and/or $ \check U $ become large, the perturbation $ E $ can become large.  
Such element growth can happen particularly if large elements result from the computation $  l_{21} := a_{21} / \alpha_{11} $ .  That happens if some of the elements in $ a_{21} $ are large in magnitude relative to the magnitude of $ \alpha_{11} $. To mitigate this, row swapping (pivoting) is employed so that $ \alpha_{11} $ becomes greater than or equal to, in magnitude, any element in $ a_{21} $, and as a result $ l_{21} $ has elements bounded in magnitude by one.  
The resulting operation is known as LU factorization with (row) pivoting ({\sc LUpiv}). An unblocked  right-looking algorithm is given in Fig.~\ref{fig:LUpiv5}.

\begin{figure}[t!]

\input LUpiv_unb_var5

\caption{Unblocked Variant 5 (right-looking algorithm) for computing {\sc LUpiv}.  
$\mbox{\sc iamax}( x )$ returns the index, $\pi$ of an element with maximal magnitude.
$ P( \pi ) $ denotes the matrix that swaps the top row with the row indexed with $ \pi $.}

\label{fig:LUpiv5}
\end{figure}

\subsection{Preliminaries}
\label{sec:preliminaries}

\NoShow{
The problem with the algorithms for LUnopiv is that they could result in the computation of very large (in magnitude) elements in $ l_{21} $ if $ \alpha_{11} $ is very small relative to one of more elements in $ a_{21} $.  This, in turn, can lead to {\em element growth} in updated elements in $ A_{22}  $, which can result in numerical instability: unnecessary error in the resulting factors and the solution of $ A x = b $ if the factors are used to solve this system.
}

Some key results regarding permutations and their actions on  matrices play are required to derive algorithms for {\sc LUpiv}.  

\begin{definition}[\sc iamax]
Given vector $ x $, 
$
\mbox{\sc iamax}(x)
$
returns  the index of an element in $ x $ with largest magnitude%
\footnote{
In our discussion, indexing starts at zero.}.
\end{definition}

\begin{definition}[Primitive permutation]
\label{def:perm_vector}
Given nonnegative integer $ \pi $, the matrix $ P( \pi ) $ is the (primitive) permutation matrix of appropriate size that, when applied to a vector $ x $ of length $m > \pi$, swaps the top element, $ \chi_0 $, with the element indexed by $ \pi $, $ \chi_{\pi} $:
\[
P( \pi ) = 
\left\{
\begin{array}{cl}
I & \mbox{if $ \pi = 0 $} \\
\left( \begin{array}{c |c | c | c}
0 & 0 & 1 & 0 \\ \hline
0 & I_{\pi-1} & 0 & 0 \\ \hline
1 & 0 & 0 & 0 \\ \hline
0 & 0 & 0 & I_{m-\pi-1}.
\end{array}
\right) & \mbox{otherwise,}
\end{array}
\right.
\]
where $ I_k $ is a $ k \times k $ identity matrix and $ 0 $ equals a submatrix (or subvector) of all zeroes of appropriate size.
\end{definition}
Importantly, applying $ P( \pi ) $ to an $ m \times n $ matrix $ A $ from the left swaps the top row with the row indexed with $ \pi $.  From the context we know that in this case $ P( \pi ) $ is $ m \times m $.

A classic result about permutation matrices:
\begin{theorem}
For any permutation matrix $ P $, its transpose equals its inverse:
$
P^{-1} = P^T$.
\end{theorem}
An immediate consequence is
\begin{corollary}
Let $ P( \pi ) $ be as defined in \ref{def:perm_vector}.  Then
$ P( \pi )^{-1} = P( \pi )^T = P( \pi ) $.
\end{corollary}
This captures that the way to undo the swapping of two rows of a matrix is to swap them again.

In order to be able to state the postcondition for {\sc LUpiv} and derive blocked algorithms, we need to define and manipulate permutations that are the aggregate of primitive permutations.
\begin{definition}
We call a vector {\renewcommand{\arraystretch}{0.5} $ p = \left( \begin{array}{c}
\pi_0 \\
\vdots \\
\pi_{n-1} 
\end{array}
\right) $} a permutation vector if  each $ \pi_i \in \{ 0, \ldots, m-i-1 \} $.
Here $ m \ge n $ is the row size of the matrix to which the permutations are applied.
\end{definition}
The vector $ p $ produced by the algorithm  in Fig.~\ref{fig:LUpiv5} is a permutation vector.
Associated with a permutation vector is the permutation matrix $ P( p ) $ that applies the permutations encoded in the vector $ p $:
\begin{definition}
\label{def:perm_matrix}
Given permutation vector $ p $ of size $ n $, 
$
P( p ) = 
\left( \begin{array}{c | c}
I_{n-1} & 0 \\ \hline
0 & P( \pi_{n-1} )
\end{array}
\right)
\cdots 
\left( \begin{array}{c | c}
1 & 0 \\ \hline
0 & P( \pi_1 )
\end{array}
\right)
P( \pi_0 )$.
\end{definition}

We will need to be able to apply permutations defined with partitioned permutation vectors.  The following theorem captures that to apply all permutations in $ p $, one can apply the first batch (given by $ p_T $) and then the second batch (given by $ p_B $).  Undoing these permutations means first undoing the second batch and then the undoing the first batch.
\begin{theorem}
\label{thm:part_perm}
Partition permutation vector $ 
p = \left( \begin{array}{c}
p_T  \\ \hline
p_B 
\end{array}
\right)$.
Then
\[
P( p ) = 
\left( \begin{array}{c | c}
I & 0 \\ \hline
0 & P( p_B ) 
\end{array}
\right)
P( p_T ) 
\quad
\mbox{and}
\quad
P( p )^{-1} =
P( p_T )^{-1}
\left( \begin{array}{c | c}
I & 0 \\ \hline
0 & P( p_B )
\end{array}
\right)^{-1}.
\]
\end{theorem}

A final corollary will become instrumental as we relate the state of variables before the update (in Step~6) to the state after the update (in Step~7), in order to determine updates (in Step~8).
\begin{corollary}
\label{cor:P}
    Partition permutation vector $ 
p = \left( \begin{array}{c}
p_1  \\ \hline
p_2  
\end{array}
\right)$.
Then
\NoShow{
\[
P( \left( \begin{array}{c}
p_1  \\ \hline
p_2  
\end{array}
\right) )^{-1} =
P( \pi_1 )^{-1}
\left( \begin{array}{c | c}
1 & 0 \\ \hline
0 & P( p_2 ) 
\end{array}
\right)^{-1}
=
P( \pi_1 )
\left( \begin{array}{c | c}
1 & 0 \\ \hline
0 & P( p_2 ) 
\end{array}
\right)^{-1}
 .
\]
and
}
\[
\left( \begin{array}{c | c}
I & 0 \\ \hline
0 & P( p_2 ) 
\end{array}
\right)^{-1}
P( \left( \begin{array}{c}
p_1  \\ \hline
p_2  
\end{array}
\right) ) = P( p_1 ) 
\quad
\mbox{and}
\quad
P( p_1 ) P( \left( \begin{array}{c}
p_1  \\ \hline
p_2  
\end{array}
\right) )^{-1} =
\left( \begin{array}{c | c}
I & 0 \\ \hline
0 & P( p_2 ) 
\end{array}
\right)^{-1}
.\]
\end{corollary}
A special case of this is when $ p_1 = \pi_1 $, a scalar:
$
P( \pi_1 ) P( \left( \begin{array}{c}
\pi_1  \\ \hline
p_2  
\end{array}
\right) )^{-1} =
\left( \begin{array}{c | c}
1 & 0 \\ \hline
0 & P( p_2 ) 
\end{array}
\right)^{-1}$.
This  provides  details for the unblocked right-looking algorithm (Variant 5) for {\sc LUpiv} given  in Fig.~\ref{fig:LUpiv5}.

\subsection{Specifying LUpiv}
\label{sec:postcondition}

The algorithm outlined at the beginning of Section~\ref{sec:LUpiv} swaps the entire row of $ A $ in each iteration, even the part that has been overwritten by elements of $ L $ in the previous iterations.
This means that, when finished,  the LU factorization is that of a matrix that equals the original matrix but with its rows permuted exactly as dictated by the pivots encountered in the LU factorization with pivoting:
$ P(p) A = L U $, where $ p $ is the permutation vector.
This insight makes it easy to compute the solution to $ A x = b $:
Since $ P(p) A x = P(p) b $, we find that $ L U x = P(p) b $. Hence,  $ x $ can be computed from the permuted vector $ b $ via two triangular solves.
We formalize this insight into a postcondition for {\sc LUpiv}.

The following theorem 
states the relationship between the matrix $ \widehat A $ that is input to that algorithm, the resulting permutation vector $ p $, and the computed factors $ L $ and $ U $.
\begin{theorem}
\label{thm:postcondition}
Let $ \widehat A $ be an $  m \times n  $ matrix, with $ n \leq m $, and assume that the algorithm in Fig.~\ref{fig:LUpiv5} is executed starting with $ A = \widehat A $.  If the algorithm completes without encountering a zero $ \upsilon_{11} $,  then the algorithm overwrites
$ A $ with $ L $ and $ U $ so that
\begin{equation}
 \label{eqn:postcond_LUpiv}
A = L \backslash U
 \wedge P( p ) \widehat A = L U \wedge  \vert L \vert \leq J .
\end{equation}
Here
\begin{itemize}
\item
$ L $ is unit lower trapezoidal and $ U $ is upper triangular with nonzero diagonal elements (meaning it is nonsingular).
\item 
$ L \backslash U$ denotes the matrix comprised of $ U $ on and above its diagonal, and the strictly lower triangular part of $ L $ below its diagonal (meaning the ones on the diagonal of  $ L $ are implicitly stored).
\item
$ p $ is a pivot vector of size $ n $.
\NoShow{
    \item
     $  \future{ A} $ equals the original matrix $ \widehat A $, but with the rows permuted consistent with the encountered row swaps.
     }
     \NoShow{
    \item
    $ \vert L \vert $ equals the matrix obtained from $ L $ by taking the element-wise absolute value.
    }
    \item
    $J $ is the matrix of all ones of appropriate size.
    \NoShow{
    \item
    $ X \leq Y $ is {\em true} iff each element of matrix $ X $ is less then or equal to the corresponding element of $ Y $.
    }
\end{itemize}
\end{theorem}
The term $ \vert L \vert \leq J $ captures that all multipliers
(values computed in $ l_{21} $)
are bounded in magnitude by one.  
It forces row swapping and mitigates element growth.
Showing that the algorithm  in Fig.~\ref{fig:LUpiv5} can be derived to  be correct  from the precondition and the postcondition proves this theorem. 
Equation~(\ref{eqn:precond_LUpiv}) is the postcondition for {\sc LUpiv}.

It is well known that the algorithm completes in the state described in Theorem~\ref{thm:postcondition} if and only if $ \widehat A $ has linearly independent columns.
However, we are going to state the precondition as
\begin{equation}
    \label{eqn:precond_LUpiv}
A = \widehat A
 \wedge P( \future{ p }) \widehat A = \future{ L }\future{  U} \wedge  \vert \future{ L }\vert \leq J .
\end{equation}
Here $ \widehat A $ equals the original content of $ A $, and  $ \future{ p} $, $ \future{ L} $, and $ \future{ U} $ refer to the $ p $, $ L $, and $ U $ in the postcondition, but the tilda capture that those quantities are known/assumed to exist, but have not yet been computed (with the usual constraints on $ p$, $ L $, and $ U $).  

With these insights and notation, we can now define {\sc LUpiv}:
\begin{definition}
\label{def:LUpiv}
    $ [ A, p ] := \mbox{\sc LUpiv}( A ) $ is the operation that, when started in a state where (\ref{eqn:precond_LUpiv}) is {\em true}, will complete in a state where (\ref{eqn:postcond_LUpiv}) is {\em true}.
\end{definition}
Using Hoare logic~\cite{10.1145/363235.363259}, this can be expressed as the Hoare triple
   \[ \{ A = \widehat A
 \wedge P( \future{ p} ) \widehat A = \future{ L} \future{  U} \wedge  \vert \future{ L} \vert \leq J  \} 
 ~~~~~~
 [ A, p ] := \mbox{\sc LUpiv}( A )  ~~~~~~
 \{ A = L \backslash U
 \wedge P( p ) \widehat A = L U \wedge  \vert L \vert \leq J  \}. 
 \]

The special case where $ A $ has only one column ($ n = 1 $) is addressed by 
\begin{theorem}
\label{thm:iamax}
    $ [ \left( \begin{array}{c}
    \alpha_{1} \\ \hline
    a_{2}
    \end{array}
    \right), \pi ] :=  \mbox{\sc LUpiv}( \left( \begin{array}{c}
    \alpha_{1} \\ \hline
    a_{2}
    \end{array}
    \right) ) $ can be implemented by the commands
    \[
\begin{array}{l}
    \pi := \mbox{\sc iamax}( \left( \begin{array}{c}
    \alpha_1 \\ \hline
    a_2
    \end{array} \right)  ) \\
    \left( \begin{array}{c}
    \alpha_1 \\ \hline
    a_2
    \end{array} \right) 
    := P( \pi )
    \left( \begin{array}{c}
    \alpha_1 \\ \hline
    a_2
    \end{array} \right) \\
    a_2 := a_2 / \alpha_1.
    \end{array}
    \]  
\end{theorem}
Here, the definition of {\sc LUpiv} implies the division is well defined since $ \upsilon_1 $ is assumed to be nonzero and hence the column must contain a  nonzero  element.

\begin{proof}
    See Appendix~\ref{appendix:proof}. 
\end{proof}

\begin{theorem}
    An $ m \times n $ matrix $ A $ with $ n \leq m $  has an LU factorization with partial pivoting, where all diagonal elements of $ U $ are nonzero, if and only if its columns are linearly independent.
\end{theorem}
This follows immediately from the fact that $ P(p)  A = L U $.  It also tells us when the algorithm is guaranteed to complete (in exact arithmetic), since it implies all encountered $ \alpha_{11} $ are nonzero.

\subsection{Deriving the PME}

We now  turn  postconditions
into PMEs.  

\subsubsection{PME I}

Starting with
(\ref{eqn:postcond_LUpiv}),
$
A = L \backslash U
 \wedge
 P(p) \widehat A = 
L U \wedge  \vert L \vert \leq J 
$,
and partitioning the various matrices and vector $ p $ 
yields
\begin{eqnarray*}
\FlaTwoByTwo
  { A_{TL} }{ A_{TR} } 
  { A_{BL} }{ A_{BR} }
  =
  \FlaTwoByTwo
  { L \backslash U_{TL} }{ U_{TR} } 
  { L_{BL} }{ L \backslash U_{BR} }
  &\wedge &
  P( \left( \begin{array}{c}
p_T  \\ \whline
p_B 
\end{array}
\right)
)
   \FlaTwoByTwo
  {\widehat A_{TL}}{\widehat A_{TR}}
  {\widehat A_{BL}}{\widehat A_{BR}} 
 = 
\FlaTwoByTwo
  { L_{TL} }{ 0 } 
  { L_{BL} }{ L_{BR} }
  \FlaTwoByTwo
  { U_{TL} }{ U_{TR} } 
  { 0 }{ U_{BR} } \\
  &\wedge  & 
\FlaTwoByTwo
  { \vert L_{TL} \vert }{ 0 } 
  { \vert L_{BL} \vert }{ \vert L_{BR} \vert }
 \leq 
 \FlaTwoByTwo
  { J }{ J } 
  { J }{ J }
  .
\end{eqnarray*}
\NoShow{
\begin{equation}
    \label{eqn:Atilde}
    \begin{array}{@{}l}
\FlaTwoByTwo
  {\future{ A_{TL}}}{\future{ A_{TR}}}
  {\future{ A_{BL}}}{\future{ A_{BR}}}
  =
P( \FlaTwoByOne{ p_T }{ p_B } ) \FlaTwoByTwo
  {\widehat A_{TL}}{\widehat A_{TR}}
  {\widehat A_{BL}}{\widehat A_{BR}} \\
  ~~~~
\wedge
\FlaTwoByTwo
  {\future{ A_{TL}}}{\future{ A_{TR}}}
  {\future{ A_{BL}}}{\future{ A_{BR}}}
  =
\FlaTwoByTwo{ L_{TL} }{0} {L_{BL} }{L_{BR}}
\FlaTwoByTwo{ U_{TL} }{U_{TR}}{0}{U_{BR}}
\end{array}.
\end{equation}
}
There are a number of useful equivalent ways of expressing the second and third term.  For example,
\begin{eqnarray}
\lefteqn{
%
    \label{eqn:constraint1}
P( p_T )
\FlaTwoByTwo
  {\widehat A_{TL}}{\widehat A_{TR}}
  {\widehat A_{BL}}{\widehat A_{BR}} 
=
\FlaTwoByTwo{ L_{TL} U_{TL} }{ L_{TL} U_{TR}} {   P(p_B)^{-1} L_{BL}  U_{TL} }{  P(p_B)^{-1} L_{BL} U_{TR} +  P(p_B)^{-1} L_{BR}  U_{BR} } 
} \hspace{2in}
    \\  \label{eqn:constraint2}
& &  \wedge
\left( \begin{array}{c I c}
  { \vert L_{TL} \vert } & { 0 } \\ \whline
  { \vert P( p_{B} )^{-1} L_{BL} \vert }& { \vert  P( p_{B} )^{-1}
  L_{BR} \vert }
    \end{array}
    \right)
    \leq 
\left( \begin{array}{c I c}
  { J } & { J } \\ \whline
  { J } & { J }
  \end{array}
\right).
\end{eqnarray}
\NoShow{or
\begin{eqnarray}
\lefteqn{
    \label{eqn:constraint3}
\FlaTwoByTwo
  {\widehat A_{TL}}{\widehat A_{TR}}
  {\widehat A_{BL}}{\widehat A_{BR}} 
=
\FlaOneByTwo{ 
P( p_T )^{-1}  \FlaTwoByOne{
L_{TL} U_{TL} }{   P(p_B)^{-1} L_{BL}  U_{TL} }}
{
P( \FlaTwoByOne{p_T}{p_B} )^{-1} 
\FlaTwoByOne{ L_{TL} U_{TR}} {  P(p_B)^{-1} L_{BL} U_{TR} +  P(p_B)^{-1} L_{BR}  U_{BR} }}
} \hspace{2in}
    \\  \label{eqn:constraint2}
& &  \wedge
\left( \begin{array}{c I c}
  { \vert L_{TL} \vert } & { 0 } \\ \whline
  { \vert P( p_{B} )^{-1} L_{BL} \vert }& { \vert  P( p_{B} )^{-1}
  L_{BR} \vert }
    \end{array}
    \right)
    \leq 
\left( \begin{array}{c I c}
  { J } & { J } \\ \whline
  { J } & { J }
  \end{array}
\right)
\end{eqnarray}
}
Here $ P( p_B )^{-1} L_{BL} $ and $ P( p_B )^{-1} L_{BR} $ capture the final $ L_{BL} $ and $ L_{BR} $ but with the rows not yet permuted according to $ p_B $.  We will refer to  (\ref{eqn:constraint1}) and 
(\ref{eqn:constraint2}) as the {\em constraints} for PME~I.

\subsubsection{PME II}

An alternative way of stating the postcondition is
$
A = L \backslash U
 \wedge
 \widetilde A = P( p ) \widehat A \wedge 
 \widetilde A = 
L U \wedge  \vert L \vert \leq J 
$,
where $ \widetilde A $ 
equals $ \widehat A $, but with the rows permuted according to $ p $ (meaning it won't be known until after the computation has completed.)
Partitioning the various matrices and vector $ p $ 
yields
\begin{eqnarray*}
\lefteqn{
\FlaTwoByTwo
  { A_{TL} }{ A_{TR} } 
  { A_{BL} }{ A_{BR} }
  =
  \FlaTwoByTwo
  { L \backslash U_{TL} }{ U_{TR} } 
  { L_{BL} }{ L \backslash U_{BR} }
  \wedge 
  \FlaTwoByTwo
  { \widetilde A_{TL} }{ \widetilde A_{TR} } 
  { \widetilde A_{BL} }{ \widetilde A_{BR} }
  =
  P( \left( \begin{array}{c}
p_T  \\ \whline
p_B 
\end{array}
\right)
)
\FlaTwoByTwo
  { \widehat A_{TL} }{ \widehat A_{TR} } 
  { \widehat A_{BL} }{ \widehat A_{BR} } } \\
  &\wedge&
   \FlaTwoByTwo
  {\widetilde A_{TL}}{\widetilde A_{TR}}
  {\widetilde A_{BL}}{\widetilde A_{BR}} 
 = 
\FlaTwoByTwo
  { L_{TL} }{ 0 } 
  { L_{BL} }{ L_{BR} }
  \FlaTwoByTwo
  { U_{TL} }{ U_{TR} } 
  { 0 }{ U_{BR} } 
 \wedge
\FlaTwoByTwo
  { \vert L_{TL} \vert }{ 0 } 
  { \vert L_{BL} \vert }{ \vert L_{BR} \vert }
 \leq 
 \FlaTwoByTwo
  { J }{ J } 
  { J }{ J }
  .
\end{eqnarray*}
The second through fourth terms can be manipulated into
\begin{eqnarray*}
\lefteqn{
\FlaTwoByTwo
  { A_{TL} }{ A_{TR} } 
  { A_{BL} }{ A_{BR} }
  =
  \FlaTwoByTwo
  { L \backslash U_{TL} }{ U_{TR} } 
  { L_{BL} }{ L \backslash U_{BR} }
  \wedge 
  \FlaTwoByTwo
  { \widetilde A_{TL} }{ \widetilde A_{TR} } 
  { P( p_B )^{-1} \widetilde A_{BL} }{ P( p_B )^{-1} \widetilde A_{BR} }
  =
  P( 
p_T  
)
\FlaTwoByTwo
  { \widehat A_{TL} }{ \widehat A_{TR} } 
  { \widehat A_{BL} }{ \widehat A_{BR} } } \\
  &\wedge&
   \FlaTwoByTwo
  {\widetilde A_{TL}}{\widetilde A_{TR}}
  {P( p_B )^{-1} \widetilde A_{BL}}{P( p_B )^{-1} \widetilde A_{BR} - P( p_B )^{-1} L_{BL} U_{TR}} 
 = 
\FlaTwoByTwo
  { L_{TL} U_{TL} }{ L_{TL} L_{TR} } 
  { P( p_B )^{-1} L_{BL} U_{TL} }{ P( p_B )^{-1} L_{BR} U_{BR} }
    \\
  &\wedge  & 
\FlaTwoByTwo
  { \vert L_{TL} \vert }{ 0 } 
  { \vert P( p_B )^{-1} L_{BL} \vert }{ \vert P( p_B )^{-1} L_{BR} \vert }
 \leq 
 \FlaTwoByTwo
  { J }{ J } 
  { J }{ J }
  ,
\end{eqnarray*}
which we will refer to as the constraints for PME~II.

In this paper, we favor invariants derived from PME I, since then the first term in the invariant, which captures the current contents of $ A $, encodes most of the information needed for the derivation, making the encountered formulae more concise.

\subsection{Deriving loop invariants from the PMEs}
\label{sec:LUpiv-invs}

\NoShow{
{\color{blue}{
\begin{equation}
    \label{eqn1}
    \begin{array}{@{}l}
\FlaTwoByTwo
  {I}{0}
  {0}{P(p_B)}
  P(p_T)
\FlaTwoByTwo
  {\widehat A_{TL}}{\widehat A_{TR}}
  {\widehat A_{BL}}{\widehat A_{BR}}  
  =
  \FlaTwoByTwo{ \future{ L_{TL}} \future{ U_{TL}} }{ \future{ L_{TL} }\future{ U_{TR}}} { \future{ L_{BL}} \future{ U_{TL}} }{  \future{ L_{BL} }\future{ U_{TR} }+ \future{ L_{BR}} \future{ U_{BR} } }.  
\end{array}
\end{equation}
\[
    \begin{array}{@{}l}
\FlaTwoByTwo{A_{TL}}{A_{TR}}
            {A_{BL}}{A_{BR}}
       =
P(p_T)       
\FlaTwoByTwo
  {\widehat A_{TL}}{\widehat A_{TR}}
  {\widehat A_{BL}}{\widehat A_{BR}}
  =
  \FlaTwoByTwo{ \future{ L_{TL}} \future{ U_{TL} }}{ \future{ L_{TL}} \future{ U_{TR}} }{ \future{ L_{BL}} \future{ U_{TL} }}{  \future{ L_{BL}} \future{ U_{TR}} + \future{ L_{BR}} \future{ U_{BR}  }}.  
\end{array}
\]
{\color{red}{There is no requirement that $p_T$ is computed before $p_B$.}}
The PME is n expression that describes how the original inputs (denoted with hats) are related to the output values (denoted with tilde). More importantly, the PME captures the computations that are required to compute the outputs from the input matrices. 

A loop invariant is an expression that describes the state of the computations at the start and end of every iteration of the loop. This means that the loop invariant must describe a partial state of computation. As such, we can 
}}
}

We hope to find algorithmic variants that require no more computation  (other than swapping rows) than the unblocked right-looking algorithm in Figure~\ref{fig:LUpiv5}.  
This dictates that in the current iteration it must be easy (cheap in terms of computation) to determine the next entry in the pivot vector.
That means that the elements in the ``current column'' (the column that gets exposed in the current iteration) must either already be up to date or must be updated in the current iteration to where the next pivot can be determined.
As a result, not all loop invariants for {\sc LUnopiv} have a natural equivalent loop invariant for {\sc LUpiv}.     
This analysis points us towards incorporating pivoting into loop invariants 3, 4, and~5.
\NoShow{
{\color{red} I think we can sharpen this  discussion.}  Starting with the old formulation for invariants, a loop invariant that closely resembles Invariant~1 from Section~\ref{sec:LUnopiv:derivation} is given by
\begin{eqnarray*}
\lefteqn{
\left( \begin{array}{c I c}
  { A_{TL} } & { A_{TR} } \\ \whline
  { A_{BL} } & { A_{BR} }
  \end{array}
  \right)
  =
  \left( \begin{array}{c I c}
  { L \backslash U_{TL} } & { \widehat A_{TR} } \\ \whline
  { \widehat A_{BL} } & { \widehat A_{BR} }
    \end{array}
    \right)
    }
    \\
& &
\wedge  ~
  P( p_T )
\left( \begin{array}{c I c}
  {\widehat A_{TL}} &{\widehat A_{TR}} \\ \whline
  {\widehat A_{BL}} &{\widehat A_{BR}}  
    \end{array}
 \right)
  =
\left( \begin{array}{c I c}
{
L_{TL} U_{TL} } & { L_{TL} U_{TR}} \\ \whline
{ P( p_B) ^{-1} L_{BL} U_{TL} } & { P( p_B) ^{-1} ( L_{BL} U_{TR} + L_{BR} U_{BR} )  }
  \end{array}
    \right)
    \\
& &
\wedge 
\left( \begin{array}{c I c}
  { \vert L_{TL} \vert } & { 0 } \\ \whline
  { \vert P( p_{B} )^{-1} L_{BL} \vert }& { \vert  P( p_{B} )^{-1} L_{BR} \vert }
    \end{array}
    \right)
    \leq 
\left( \begin{array}{c I c}
  { J } & { J } \\ \whline
  { J } & { J }
  \end{array}
\right)
    .
\end{eqnarray*}
There is a fundamental problem with this expression:
Notice that some of the rows of $ \widehat A_{TL} $ should have been swapped into the bottom-left quadrant, and would thus be lost.
This can be fixed by instead taking for the first term in the invariant as
\[
\left( \begin{array}{c I c}
  { A_{TL} } & { A_{TR} } \\ \whline
  { A_{BL} } & { A_{BR} }
  \end{array}
  \right)
  =
  \left( \begin{array}{c I c}
  { L \backslash U_{TL} } & { \widehat A_{TR} } \\ \whline
  { P( p_B) ^{-1} L_{BL} U_{TL} } & { \widehat A_{BR} }
    \end{array}
    \right)
\]
since 
$
P( p_B )^{-1}
  L_{BL} U_{TR}
  $
  equals $ \widehat A_{BL} $ except with the appropriate rows swapped in the required way.
Still, in general, the pivot vector $ p_T $ cannot have been determined without also having  computed with $ A_{BL} $.  Thus, this invariant  leads to an algorithm that likely performs computation that is then discarded (unless no pivoting happens), which would be inefficient.
A third concern is that when discussing blocked algorithms for LUpiv we will see that we will want to factor matrices $ A \in \mathbb{R}^{m \times n} $ where $ n \leq m $.  Variants~1 and~2  for LUnopiv do not complete in the correct state if $ n < m $ and hence they cannot be expected to do so for LUpiv.  (For LUnopiv, this can be fixed by performing a post process to update the remaining part of the matrix.)

This suggests that an examination of the dependencies between different parts of $ A $ and $ p $ must be performed to determine viable loop invariants.  Relating this back to the right-looking LUpiv algorithm, notice that in a current iteration the index $ \pi_1 $ must be computed.  This means that either $ \left( \begin{array}{c}
\alpha_{11} \\ \hline
a_{21} 
\end{array} \right) $ must already have  been updated in previous iterations (as it is for the right-looking algorithm) {\em or} it must be easy to compute in the current iteration.  Thus, 
$ \left( \begin{array}{c} 
a_{10}^T \\ \hline
A_{20} 
\end{array} \right) $ must already have been updated to where they can be used to compute  $ \left( \begin{array}{c}
\alpha_{11} \\ \hline
a_{21} 
\end{array} \right) $
to the point where 
$ \pi_1 $ can be determined.
That, in turn, means that any loop invariant that is a natural extention of LUnopiv Invariants~1 or~2 are not in the running.
This informal reasoning yields the following invariants for LUpiv corresponding to a subset of the invariants for LUnopiv.
} 

\paragraph*{Invariant~3piv-a.} 

Regardless of whether one starts with PME~I or~II, corresponding to Invariant 3 in Section~\ref{sec:LUnopiv:derivation} we find 
Invariant~3piv-a:
\begin{equation}
    \label{eqn:inv_3piv-a}
\begin{array}{l}
\left( \begin{array}{c I c}
  { A_{TL} } & { A_{TR} } \\ \whline
  { A_{BL} } & { A_{BR} }
  \end{array}
  \right)
  =
  \left( \begin{array}{c I c}
  { L \backslash U_{TL} } & { \widehat A_{TR} } \\ \whline
  { P( \future{ p_B} )^{-1} L_{BL} } & { \widehat A_{BR}  }
    \end{array}
    \right)
\wedge 
\mbox{constraints} .
\end{array}
\end{equation}
This indicates that only the left quadrants have been updated with the final result, as in Invariant~3 for LUnopiv, except that the rows of $ L_{BL} $ must still be swapped consistent with ``future'' pivots.  The right quadrants have not yet been updated.
This can also be expressed as
\[
\begin{array}{l}
\left( \begin{array}{c I c}
  { A_{TL} } & { A_{TR} } \\ \whline
  { A_{BL} } & { A_{BR} }
  \end{array}
  \right)
  =
  \left( \begin{array}{c I c} 
  \left( \begin{array}{c}
  L \backslash U_{TL}  \\ \whline
  P( \future{ p_B} )^{-1} L_{BL} 
  \end{array} \right){  } & 
  P( 
  \left( \begin{array}{c}
  p_T \\ \whline
  \future{ p_B}
  \end{array} \right)  )^{-1} \left( \begin{array}{c}
  L_{TL}  \future{ U_{TR}}  \\ \whline
   L_{BL} \future{ U_{TR}} + \future{ L_{BR} }\future{ U_{BR}   }
  \end{array} \right) 
    \end{array}
    \right)
\wedge 
\mbox{constraints} .
\end{array}
\]

\paragraph*{Invariant~3piv-b.} 

A slight variation on Invariant 3piv-a is given by
\[
\begin{array}{l}
\left( \begin{array}{c I c}
  { A_{TL} } & { A_{TR} } \\ \whline
  { A_{BL} } & { A_{BR} }
  \end{array}
  \right)
  =
  \left( \begin{array}{c I c} 
  \left( \begin{array}{c}
  L \backslash U_{TL}  \\ \whline
  P( \future{ p_B} )^{-1} L_{BL} 
  \end{array} \right){  } & 
  P( 
   p_T   ) \left( \begin{array}{c}
  \widehat A_{TR}  \\ \whline
   \widehat A_{BR} 
  \end{array} \right) 
    \end{array}
    \right)
\wedge 
\mbox{cconstraints} .
\end{array}
\]
It differs from Invariant~3-piv-a in that row swaps have also been applied to right quadrants.
Notice that involving the original contents of $ A $, $ \widehat A $, makes this expression awkward.

Using PME~I, 
it can instead be expressed as
\[
\begin{array}{l}
\left( \begin{array}{c I c}
  { A_{TL} } & { A_{TR} } \\ \whline
  { A_{BL} } & { A_{BR} }
  \end{array}
  \right)
  =
  \left( \begin{array}{c I c}
  { L \backslash U_{TL} } & { L_{TL} \future{ U_{TR}} } \\ \whline
  { P( \future{ p_B })^{-1} L_{BL} } & { P( \future{ p_B}) ^{-1} L_{BL} \future{ U_{TR}} + P( \future{ p_B}) ^{-1} \future{ L_{BR}} \future{ U_{BR} }  }
    \end{array}
    \right)
\wedge 
\mbox{constraints} 
\end{array}
\]
while PME~II yields
\[
\begin{array}{l}
\left( \begin{array}{c I c}
  { A_{TL} } & { A_{TR} } \\ \whline
  { A_{BL} } & { A_{BR} }
  \end{array}
  \right)
  =
  \left( \begin{array}{c I c}
  { L \backslash U_{TL} } & \future{ A_{TR}}  \\ \whline
  { P( \future{ p_B })^{-1} L_{BL} } & { P( \future{ p_B})^{-1} \future{ A_{BR}}  }
    \end{array}
    \right)
\wedge 
\mbox{constraints} .
 
\end{array}
\]

\paragraph*{Invariant~4piv.} 

Corresponding to Invariant 4 in Section~\ref{sec:LUnopiv:derivation} PME~I yields
Invariant~4piv:
\[
\begin{array}{l}
\left( \begin{array}{c I c}
  { A_{TL} } & { A_{TR} } \\ \whline
  { A_{BL} } & { A_{BR} }
  \end{array}
  \right)
  =
  \left( \begin{array}{c I c}
  { L \backslash U_{TL} } & { U_{TR} } \\ \whline
  { P( \future{ p_B} )^{-1}  L_{BL} } & { P( \future{ p_B} ) ^{-1}  L_{BL} U_{TR} + P( \future{ p_B}) ^{-1}  \future{ L_{BR}} \future{ U_{BR}}  }
    \end{array}
    \right)
\wedge 
\mbox{constraints} ,
\end{array}
\]
while PME~II yields
\[
\begin{array}{l}
\left( \begin{array}{c I c}
  { A_{TL} } & { A_{TR} } \\ \whline
  { A_{BL} } & { A_{BR} }
  \end{array}
  \right)
  =
  \left( \begin{array}{c I c}
  { L \backslash U_{TL} } & { U_{TR} } \\ \whline
  { P( \future{ p_B} )^{-1}  L_{BL} } & { P( \future{ p_B} ) ^{-1}  \future{ A_{BR}}  }
    \end{array}
    \right)
\wedge 
\mbox{constraints} .
\end{array}
\]
Again, this invariant is difficult to  expressed in  terms of $ \widehat A_{TR} $ and 
$ \widehat A_{BR} $, since it involves  $ A_{BR} $ that results from swapping rows between $ \widehat A_{TR} $ and $ \widehat A_{BR} $.  

\paragraph*{Invariant~5piv.} 
Corresponding to Invariant 5 in Section~\ref{sec:LUnopiv:derivation}, PME~I yields  Invariant 5piv
\begin{equation}
    \label{eqn:inv5piva}
\begin{array}{l}
\left( \begin{array}{c I c}
  { A_{TL} } & { A_{TR} } \\ \whline
  { A_{BL} } & { A_{BR} }
  \end{array}
  \right)
  =
  \left( \begin{array}{c I c}
  { L \backslash U_{TL} } & { U_{TR} } \\ \whline
  { P( \future{ p_B}) ^{-1} L_{BL} } & { 
  P( \future{  p_B} )^{-1} 
  \future{ L_{BR}}\future{ U_{BR} }}
    \end{array}
    \right)
\wedge 
\mbox{constraints} 
    
\end{array}
\end{equation}
while PME~II yields
\[
\begin{array}{l}
\left( \begin{array}{c I c}
  { A_{TL} } & { A_{TR} } \\ \whline
  { A_{BL} } & { A_{BR} }
  \end{array}
  \right)
  =
  \left( \begin{array}{c I c}
  { L \backslash U_{TL} } & { U_{TR} } \\ \whline
  { P( \future{ p_B}) ^{-1} L_{BL} } & { 
  P( \future{  p_B} )^{-1} \future{ A_{BR}} -
  P( \future{  p_B} )^{-1}  L_{BL} U_{TR} }
    \end{array}
    \right)
\wedge 
\mbox{constraints} .
    
\end{array}
\]
Again,  it would difficult to  express this invariant  in  terms of $ \widehat A_{TR} $ and 
$ \widehat A_{BR} $.  

\subsection{From invariant to unblocked algorithm}
\label{sec:LUpiv:unb}

We next show how these invariants can be turned into algorithmic variants.

\subsubsection{Deriving unblocked algorithms for LUpiv}

To illustrate how unblocked algorithms can be derived, we choose Invariant~5piv as derived from PME~I, in (\ref{eqn:inv5piva}), for Step 2.
Since Steps~3-5 are routine, we focus on the predicates in Steps~6 and ~7, and how to then determine the update in Step 8.
While it is impractical to fill the worksheet due to the length of the various predicates, the reader should keep the worksheet in mind.

Repartitioning the various variables and substituting the result into the invariant yields the following for Step 6:
\begin{eqnarray}
\nonumber
~~~~~\left( \begin{array}{ c I c | c }
A_{00} &
a_{10} & A_{20} \\ \whline
&&\\[-15pt]
\left( \begin{array}{c}
a_{10}^{T} \\ \hline
A_{20}
\end{array} \right) & 
\left( \begin{array}{c}
\alpha_{11} \\ \hline
a_{21}
\end{array} \right) & 
\left( \begin{array}{c}
a_{12}^{T} \\ \hline
A_{22}
\end{array} \right)
\end{array}
\right) =
\left( \begin{array}{ c I c | c}
L \backslash U_{00} & \quad\quad\quad u_{01} \quad\quad\quad & U_{02} \\ \whline
\\[-15pt]
P( \left( \begin{array}{c}
\future{
\pi_1} \\ \hline
\future{
p_2}
\end{array} \right) )^{-1}
\left( \begin{array}{c}
l_{10}^T \\ \hline
L_{20}
\end{array}
\right)
&
\multicolumn{2}{c}
{
P( \left( \begin{array}{c}
\future{ \pi_1} \\ \hline
\future{ p_2}
\end{array} \right) )^{-1}
\left( \begin{array}{c | c}
1 & 0 \\ \hline
\future{ l_{21}} & \future{ L_{22}}
\end{array} \right)
\left( \begin{array}{c| c}
\future{ \upsilon_{11}} &
\future{ u_{12}^T }\\ \hline
0 & \future{ U_{22}}
\end{array}
\right)
}
\end{array}
\right) 
 \NoShow{
 \\
 \nonumber
& &
\wedge ~ 
\left( \begin{array}{ c I c | c}
\future{ A_{00}} & \future{ A_{01}} & \future{ A_{02}} \\ \whline
P( \left( \begin{array}{c}
\pi_1 \\ \hline
p_2
\end{array} \right) )^{-1}
\left( \begin{array}{c}
\future{ a_{10}^T} \\ \hline
\future{ A_{20}}
\end{array}
\right)
&

P( \left( \begin{array}{c}
\pi_1 \\ \hline
p_2
\end{array} \right) )^{-1}
\left( \begin{array}{c} 
\future{ \alpha_{11}} \\ \hline
\future{ a_{21} }
\end{array} \right)
&

P( \left( \begin{array}{c}
\pi_1 \\ \hline
p_2
\end{array} \right) )^{-1}
\left( \begin{array}{c}
\future{ a_{12}^T} \\ \hline
\future{ A_{22} }
\end{array} \right)
\end{array}
\right) \\
\nonumber
& & ~~~~~~~~~~~~~~ =
P( p_0 )
\left( \begin{array}{ c I c | c}
\widehat A_{00} & \widehat A_{01} & \widehat A_{02} \\ \whline
\left( \begin{array}{c}
\widehat a_{10}^T \\ \hline
\widehat A_{20}
\end{array}
\right)
&

\left( \begin{array}{c} 
\widehat \alpha_{11} \\ \hline
\widehat a_{21} 
\end{array} \right)
&
\left( \begin{array}{c}
\widehat a_{12}^T \\ \hline
\widehat A_{22} 
\end{array} \right)
\end{array}
\right)
\NoShow{
\future{ A} = P( p ) \widehat A}  \\ 
\nonumber
& & \wedge
  \left( \begin{array}{ c I c | c}
\future{ A_{00}} & \future{ a_{01} }& \future{ A_{02}} \\ \whline
P( \left( \begin{array}{c}
\pi_1 \\ \hline
p_2
\end{array} \right) )^{-1}
\left( \begin{array}{c}
\future{ a_{10}^T} \\
\future{ A_{20}}
\end{array}
\right)
&
&
\end{array}
\right)
=
  \left( \begin{array}{ c I c | c}
L_{00} U_{00} & L_{00} u_{01} & L_{00} U_{02} \\ \whline
 P( \left( \begin{array}{c}
\pi_1 \\ \hline
p_2
\end{array} \right) )^{-1}
\left( \begin{array}{c}
l_{10}^T \\ \hline
L_{20}
\end{array}
\right) U_{00}
&
&
\end{array}
\right)
} 
  \\
 \nonumber
\wedge 
\mbox{~(\ref{eqn:constraint1})}\wedge
 \left( \begin{array}{ c  | c }
\vert L_{00} \vert & 0 \\ \whline
&\\[-15pt]
\left\vert
P( \left( \begin{array}{c}
\future{ \pi_1} \\ \hline
\future{ p_2}
\end{array} \right) )^{-1}
\left( \begin{array}{c}
l_{10}^T \\ \hline
L_{20}
\end{array}
\right) \right\vert
&
\left\vert
P( \left( \begin{array}{c}
\future{ \pi_1 }\\ \hline
\future{ p_2}
\end{array} \right) )^{-1}
\left( \begin{array}{c|c}
1 & 0  \\ \hline
\future{ l_{21} }& 
\future{ L_{22}}
\end{array}
\right) \right\vert
\end{array}
\right)
\leq
\left( \begin{array}{c|c}
J & J \\ \whline
&\\[-15pt]
\left( \begin{array}{c}
j^T \\ \hline
J
\end{array} \right)
&
\left( \begin{array}{c|c}
1 & j^T \\ \hline 
j & J
\end{array} \right)
\end{array}
\right)
.
\end{eqnarray}
or
{
\setlength{\arraycolsep}{2pt}
\begin{eqnarray}
\nonumber
\lefteqn{
\left( \begin{array}{ c I c | c }
A_{00} &
a_{10} & A_{20}
\\ \whline
&&\\[-15pt]
\left( \begin{array}{c}
a_{10}^{T} \\ \hline
A_{20}
\end{array} \right) & 
\left( \begin{array}{c}
\alpha_{11} \\ \hline
a_{21}
\end{array} \right) & 
\left( \begin{array}{c}
a_{12}^{T} \\ \hline
A_{22}
\end{array} \right) 
\end{array}
\right)
=  
\left( \begin{array}{ c I c | c}
L \backslash U_{00} & u_{01} & U_{02} \\ \whline
&&\\[-15pt]
P( \left( \begin{array}{c}
\future{ \pi_1} \\ \hline
\future{ p_2}
\end{array} \right) )^{-1}
\left( \begin{array}{c}
l_{10}^T \\ \hline
L_{20}
\end{array}
\right)
&
P( \left( \begin{array}{c}
\future{ \pi_1} \\ \hline
\future{ p_2}
\end{array} \right) )^{-1}
\upsilon_{11}
\left( \begin{array}{c}
1  \\ \hline
\future{  l_{21} }
\end{array}
\right)
&
P( \left( \begin{array}{c}
\future{ \pi_1 }\\ \hline
\future{  p_2}
\end{array} \right) )^{-1}
\left( \begin{array}{c}
\future{ u_{12}^T} \\ \hline
\future{ l_{21}} \future{ u_{12}^T} + \future{ L_{22}} \future{ U_{22}}
\end{array}
\right)
\end{array}
\right) } \hspace{0.5in}\\
\nonumber
& \wedge \mbox{(\ref{eqn:constraint1})}   \wedge
\left( \begin{array}{ c I c | c}
\vert L_{00}  \vert  & 0 & 0 \\ \whline
&&\\[-15pt]
\left\vert P( \left( \begin{array}{c}
\future{ \pi_1} \\ \hline
\future{ p_2}
\end{array} \right) )^{-1}
\left( \begin{array}{c}
l_{10}^T \\ \hline
L_{20}
\end{array}
\right)
\right\vert
&
\left\vert P( \left( \begin{array}{c}
\future{ \pi_1} \\ \hline
\future{ p_2}
\end{array} \right) )^{-1}
\left( \begin{array}{c}
1  \\ \hline
\future{ l_{21} }
\end{array}
\right)
\right\vert
&
\left\vert 
P( \left( \begin{array}{c}
\future{ \pi_1} \\ \hline
\future{ p_2}
\end{array} \right) )^{-1}
\left( \begin{array}{c}
0 \\ \hline
\future{ L_{22} }
\end{array}
\right)
\right\vert
\end{array}
\right)
\leq
\left( \begin{array}{c I c | c }
J & j & J \\ \whline
 j^T & 1  & j^T\\ \hline
J & j & J
\end{array}
\right)
.
\end{eqnarray}
}
Similarly, the state in Step~7 is:
{
\begin{eqnarray}
\nonumber
\lefteqn{
\left( \begin{array}{ c I c | c}
A_{00} &  a_{01} & A_{02} \\ \whline
 a_{10}^{T} &  \alpha_{11} &  a_{12}^{T} \\ \hline
A_{20} & a_{21} &  A_{22}
\end{array}
\right) 
    = \left( \begin{array}{ c I c | c}
L \backslash U_{00} & u_{01} & U_{02} \\ \whline l_{10}^T &  \upsilon_{11} &  u_{12}^T \\
\hline 
P( 
\future{ p_2} )^{-1}
 L_{20}
& 
P( 
\future{  p_2}
 )^{-1}
l_{21}  
&
P( 
{
\future{ 
p_2}
}
 )^{-1}
\future{  L_{22}}\future{ U_{22}} 
\end{array}
\right) \wedge \mbox{(\ref{eqn:constraint1})} 
}
  \\
\nonumber
& & \wedge
 \left( \begin{array}{ c I c | c }
\vert L_{00} \vert 
& 0 & 0  \\ \whline 
\vert l_{10}^T  \vert  & 1 & 0 \\ \hline
\vert P( 
p_2
 )^{-1}
L_{20} 
\vert
&
\vert  P( 
\future{  p_2} )^{-1} l_{21}  \vert
& 
\vert  P( 
\future{  p_2} )^{-1} \future{  L_{22}}  \vert
\end{array}
\right)
\leq
\left( \begin{array}{c I c | c }
J & j & J \\ \whline
 j^T & 1  & j^T\\ \hline
J & j & J
\end{array}
\right)
.
\end{eqnarray}
}%
The goal now is to reason through how to update the contents of $ A $ from what they contain in Step~6 to what they must contain in  Step~7.  
For this, we let $ A^{\mbox{\rm cur}} $ denote the state of $ A $ at Step~6.  This means, with a bit of manipulation, at Step~6 the state of $ \alpha_{11} $ and $ a_{21} $, etc., is
\[
\left( \begin{array}{c}
\alpha_{11} \\ \hline
a_{21}
\end{array} \right) 
=
\left( \begin{array}{c}
\alpha_{11}^{\mbox{\rm cur}} \\ \hline
a_{21}^{\mbox{\rm cur}}
\end{array} \right) 
\wedge
P( \future{ \pi_1} )
\left( \begin{array}{c}
\alpha_{11}^{\mbox{\rm cur}} \\ \hline
a_{21}^{\mbox{\rm cur}}
\end{array} \right) 
=
\left( \begin{array}{c}
1 \\ \hline
P( 
\future{ p_2}
)^{-1} \future{ l_{21}}
\end{array} \right)
\future{ \upsilon_{11}}
\wedge
\left\vert 
p( \pi_1 )^{-1} 
\left( \begin{array}{c}
1  \\ \hline
P( \future{ p_2} )^{-1} \future{ l_{21} }
\end{array}
\right)
\right\vert
\leq
\left( \begin{array}{c}
1  \\ \hline
 j
\end{array}
\right),
\]
where implicitly it is known that $ \future{  \upsilon_{11}} \neq 0 $.
After manipulation, at Step~7 
\[
\left( \begin{array}{c}
\alpha_{11} \\ \hline
a_{21}
\end{array} \right) 
=
\left( \begin{array}{c}
\upsilon_{11} \\ \hline
P( \future{ p_2}  )^{-1} l_{21}
\end{array} \right)
\wedge
P(  \pi_1 )
\left( \begin{array}{c}
\alpha_{11}^{\mbox{\rm cur}} \\ \hline
a_{21}^{\mbox{\rm cur}}
\end{array} \right) 
=
\left( \begin{array}{c}
1 \\ \hline
P( 
\future{ p_2}
)^{-1}  l_{21}
\end{array} \right)
 \upsilon_{11}
\wedge
\left( \begin{array}{c}
1 \\ \hline
\vert P( 
\future{ p_2 }
 )^{-1} l_{21} \vert
\end{array} \right)
\leq 
\left( \begin{array}{c}
1 \\ \hline
j 
\end{array} \right)
\]
\NoShow{ 
\[
\left( \begin{array}{c}
\alpha_{11} \\ \hline
a_{21}
\end{array} \right) 
 =
P( \left( \begin{array}{c}
\pi_1 \\ \hline
p_2
\end{array} \right) )^{-1}
\left( \begin{array}{c}
\upsilon_{11}  \\ \hline
\upsilon_{11} l_{21} 
\end{array}
\right)
\quad
\mbox{to}
\quad
\left( \begin{array}{c}
\alpha_{11} \\ \hline
a_{21}
\end{array} \right) 
= 
\left( \begin{array}{c}
\upsilon_{11} \\ \hline
P( p_2 )^{-1} l_{21}
\end{array} \right) .
\]
For this, we notice that
\begin{eqnarray*}
\lefteqn{
\left( \begin{array}{c }
 \alpha_{11}  \\ \hline
a_{21} 
\end{array}
\right) 
    = \left( \begin{array}{  c}
\upsilon_{11}  \\
\hline 
P( 
p_2
 )^{-1}
l_{21}  
\end{array}
\right)
=
P( \pi_1 )
P( \left( \begin{array}{c}
\pi_1 \\ \hline
p_2
\end{array} \right)
)^{-1}
\left( \begin{array}{c}
\upsilon_{11} \\ \hline
l_{21}
\end{array} \right)
}
\\
& & 
=
P( \pi_1 )
P( \left( \begin{array}{c}
\pi_1 \\ \hline
p_2
\end{array} \right)
)^{-1}
\left( \begin{array}{c}
\upsilon_{11} \\ \hline
l_{21}
\end{array} \right)
\left( \begin{array}{c|c}
1 & 0  \\ \hline
0 & I / \alpha_{11}^{\rm{next}} 
\end{array} \right)
=
P( \pi_1 )
\left( \begin{array}{c}
\alpha_{11} \\ \hline
a_{21}
\end{array} \right)
\left( \begin{array}{c|c}
1 & 0  \\ \hline
0 & I / \alpha_{11}^{\rm{next}} 
\end{array} \right).
\end{eqnarray*}
Manipulating this further yields the insight that we want to go from a precondition of 
\[
\left( \begin{array}{c}
\alpha_{11} \\ \hline
a_{21}
\end{array} \right) 
 =
\left( \begin{array}{c}
\alpha_{11} \\ \hline
a_{21}
\end{array} \right) 
\]
to a postcondition of
\[
\left( \begin{array}{c}
\alpha_{11} \\ \hline
a_{21}
\end{array} \right) 
 =
 \left( \begin{array}{c}
\upsilon_{11}  \\ \hline
 P( p_2)^{-1} l_{21} 
\end{array}
\right)
\wedge
P( \pi_1 )
\left( \begin{array}{c}
\alpha_{11} \\ \hline
a_{21}
\end{array} \right)
=
\upsilon_{11}
\left( \begin{array}{c}
1 \\ \hline
 P( p_2)^{-1} l_{21}
\end{array} \right)
\wedge
\left( \begin{array}{c}
1 \\ \hline
\vert l_{21} \vert
\end{array} \right)
\leq 
\left( \begin{array}{c}
1 \\ \hline
j 
\end{array} \right).
\]
}%
must hold.  Theorem~\ref{thm:iamax} tells us 
 the commands
\[
   \begin{array}{@{}l@{}}
   \pi_1 := \mbox{\sc iamax}( \left( \begin{array}{c} 
        \alpha_{11} \\ \hline
        a_{21}
        \end{array}
        \right) )
        \\
        \left( \begin{array}{c}
        \alpha_{11} \\ \hline
        a_{21}
        \end{array}
        \right) :=
        P( \pi_1 ) \left( \begin{array}{c}
        \alpha_{11} \\ \hline
        a_{21}
        \end{array}
        \right)\\
         a_{21} := a_{21} / \alpha_{11} 
         \end{array}
\]
will update $ \alpha_{11} $, $ a_{21} $, and $ \pi_ 1 $ appropriately.
Once $ \pi_1 $ is known, updating
\begin{equation}   \label{eqn:Lupiv_unb_before}
\left( \begin{array}{ c I c | c}
 a_{10}^{T} &  ~~ &  a_{12}^{T} \\ \hline
 A_{20}^{} &   &  A_{22}
\end{array}
\right) 
    = 
\left(
\begin{array}{c I c | c} 
P(
\left( \begin{array}{c}
\future{  \pi_1} \\ \hline
\future{  p_2}
\end{array}
\right)
)^{-1}
\left( \begin{array}{c}
l_{10}^T \\ \hline
L_{20}
\end{array}
\right)
&
~~
&
P(
\left( \begin{array}{c}
\future{  \pi_1 }\\ \hline
\future{  p_2}
\end{array}
\right)
)^{-1}
\left( \begin{array}{c}
\future{  u_{12}^T }\\ \hline
l_{21} \future{  u_{12}^T}  +  \future{  L_{22} } \future{ U_{22}}
\end{array}
\right)
\end{array}
\right)
\end{equation}
to
\[
\left( \begin{array}{ c I c | c}
 a_{10}^{T} &  ~~ &  a_{12}^{T} \\ \hline
 A_{20}^{} &   &  A_{22}
\end{array}
\right) 
    = \left( \begin{array}{ c I c | c}
  l_{10}^T &  ~~ &  u_{12}^T \\
\hline 
P( \future{  p_2} )^{-1}
 L_{20}
&  
&
P( 
\future{  p_2}
 )^{-1}
\future{  L_{22}}\future{ U_{22} }
\end{array}
\right)
\]
prescribes the updates
\[
\begin{array}{@{}l}
\left( \begin{array}{c I c | c}
a_{10}^T & & a_{12}^T \\ \hline
A_{22} & & A_{22} 
\end{array} \right)
:= 
P( \pi_1 )
\left( \begin{array}{c I c | c }
a_{12}^T  & & a_{12}^T \\ \hline
A_{22} & & A_{22} 
\end{array} \right) \\
A_{22} := A_{22} - a_{21} a_{12}^T,
\end{array}
\]
since (\ref{eqn:Lupiv_unb_before}) is equivalent to
\[
P( \future{  \pi_1} )
\left( \begin{array}{ c I c | c}
 a_{10}^{T} &  ~~ &  a_{12}^{T} \\ \hline
 A_{20}^{} &   &  A_{22}
\end{array}
\right) 
    = 
\left(
\begin{array}{c I c | c} 
\left( \begin{array}{c}
l_{10}^T \\ \hline
P( \future{  p_2} )^{-1} L_{20}
\end{array}
\right)
&
~~
&
\left( \begin{array}{c}
\future{  u_{12}^T} \\ \hline
( P( \future{  p_2} )^{-1} \future{  l_{21}} ) 
\future{  u_{12}^T}  +  P( \future{  p_2} )^{-1} \future{ L_{22} } \future{ U_{22}}
\end{array}
\right)
\end{array}
\right).
\]
This completes the derivation of the update for Variant 5 (right-looking) in Figures~\ref{fig:LUpiv5} and~\ref{fig:LUpiv-unb-all}.
The other variants in Figure~\ref{fig:LUpiv-unb-all} can be similarly derived from their invariants, which is left as an exercise for the reader.
\begin{figure}[tbp]
\input LUpiv_unb_all

\begin{center}
 \FlaAlgorithm  
\end{center}

\caption{Unblocked algorithms for {\sc LUpiv}.}
\label{fig:LUpiv-unb-all}
\end{figure}

Having derived algorithms for {\sc LUpiv} proves Theorem~\ref{thm:postcondition} and implies the operation {\sc LUpiv} in Definition~\ref{def:LUpiv} is well-defined and can hence be used in the derivation of blocked algorithms, next.

\NoShow{
\[
\begin{array}{@{}l}
   \pi_1 := \mbox{\sc iamax}( \left( \begin{array}{c} 
        \alpha_{11} \\ \hline
        a_{21}
        \end{array}
        \right) )
        \\
        \left( \begin{array}{c I c | c}
        a_{10}^T & \alpha_{11} & a_{12}^T \\ \hline
        A_{20} & a_{21} & A_{22}
        \end{array}
        \right) :=
        P( \pi_1 ) \left( \begin{array}{c I c | c}
        a_{10}^T & \alpha_{11} & a_{12}^T \\ \hline
        A_{20} & a_{21} & A_{22}
        \end{array}
        \right)\\
         a_{21} := a_{21} / \alpha_{11} \\
A_{22} := A_{22} - a_{21} a_{12}^T.
         \end{array}
         \]
         }

\subsection{Deriving blocked algorithms for LUpiv}

The process for deriving a blocked algorithm is similar.  We illustrate it for Invariant~3piv-a.
Again, 
Steps~3-5 are routine, and we focus on the predicates in Steps~6 and~7, and how to then determine the update in Step~8.

Repartioning the various variables and substituting the result into the invariant and minor  manipulation yields the following for Step 6:
\begin{equation}
    \label{eqn:step6}
\begin{array}{l}
\left( \begin{array}{c I c | c}
  \left( \begin{array}{c}
  A_{00} \\ \whline
  A_{10} \\ \hline
  A_{20}
  \end{array}
  \right)
  &
   \left( \begin{array}{c}
  A_{01} \\ \whline
  A_{11} \\ \hline
  A_{21}
  \end{array}
  \right)
  &
   \left( \begin{array}{c}
  A_{02} \\ \whline
  A_{12} \\ \hline
  A_{22}
  \end{array}
  \right)
  \end{array}
  \right)
  = 
  \\  \mypadding[-2pt]
  ~~~~
  \setlength{\arraycolsep}{2pt}
\left( \begin{array}{c I c | c}
  \left( \begin{array}{c}
  L \backslash U_{00} \\ \whline  \mypadding[-2pt]
P( 
  \left( \begin{array}{c}
  \future{  p_1} \\ \hline
  \future{  p_2}
  \end{array}
  \right)
  )^{-1}
  \left( \begin{array}{c}
  L_{10} \\ \hline
  L_{20}
  \end{array}
  \right)
  \end{array}
  \right)
  &
  P( 
  \left( \begin{array}{c}
  p_0 \\ \whline
  \future{  p_1 }\\ \hline
  \future{  p_2 }\\  
  \end{array}
  \right)
  )^{-1}
   \left( \begin{array}{c}
  L_{00} \future{  U_{01}} \\ \whline
  L_{10} \future{ U_{01} } + {\future{  L_{11}}\future{ U_{11}}} \\ \hline
  \future{ L_{20}} \future{  U_{01}} + \future{ L_{21} \future{ U_{11}} }
  \end{array}
  \right)
  &
  P( 
  \left( \begin{array}{c}
  p_0 \\ \whline
  \future{  p_1} \\ \hline
  \future{  p_2 }\\  
  \end{array}
  \right)
  )^{-1}
   \left( \begin{array}{c}
  L_{00} \future{  U_{02}} \\ \whline
  L_{10} \future{  U_{02}} + \future{ L_{11}} \future{ U_{12}} \\ \hline
  L_{20} \future{  U_{02}} + \future{  L_{22}} \future{ U_{22} }
  \end{array}
  \right)
  \end{array}
  \right) 
\wedge 
\mbox{constraints} .
\end{array}
\end{equation}
Similarly, the state in Step~7, after some manipulation, is:
\begin{equation}
\label{eqn:step7}
\begin{array}{l}
\left( \begin{array}{c I c | c}
  \left( \begin{array}{c}
  A_{00} \\ \whline
  A_{10} \\ \hline
  A_{20}
  \end{array}
  \right)
  &
   \left( \begin{array}{c}
  A_{01} \\ \whline
  A_{11} \\ \hline
  A_{21}
  \end{array}
  \right)
  &
   \left( \begin{array}{c}
  A_{02} \\ \whline
  A_{12} \\ \hline
  A_{22}
  \end{array}
  \right)
  \end{array}
  \right)
  = 
  \setlength{\arraycolsep}{2pt}
\left( \begin{array}{c I c | c}
  \left( \begin{array}{c}
  L \backslash U_{00} \\ \whline
    L_{10} \\ \hline
  P( 
  \future{  p_2}
  )^{-1}
  L_{20}
  \end{array}
  \right)
  &
   \left( \begin{array}{c}
U_{01} \\ \whline
 L \backslash U_{11}  \\ \hline
  P( 
  \future{  p_2}
  )^{-1}  L_{21}
  \end{array}
  \right)
  &
  P( 
  \left( \begin{array}{c}
  p_0 \\ \whline
  p_1 \\ \hline
  \future{  p_2} \\  
  \end{array}
  \right)
  )^{-1}
   \left( \begin{array}{c}
  \future{  L_{00}} \future{ U_{02}} \\ \whline
  L_{10} \future{   U_{02}} + \future{  L_{11} }\future{ U_{12} }\\ \hline
  \future{  L_{20} }\future{ U_{02}} + \future{  L_{22} \future{ U_{22}}} 
  \end{array}
  \right)
  \end{array}
  \right) 
\wedge 
\mbox{constraints} .
\end{array}
\end{equation}
From this, we now reason what updates take matrix $ A $ from the state described in Step~6 to the state described in Step~7, thus providing the update in Step~8.

From Step~6 we find that
\[
   \left( \begin{array}{c}
  A_{01} \\ \whline
  A_{11} \\ \hline
  A_{21}
  \end{array}
  \right)
  =
P( 
  \left( \begin{array}{c}
  p_0 \\ \whline
  \future{  p_1} \\ \hline
  \future{  p_2 }\\  
  \end{array}
  \right)
  )^{-1}
   \left( \begin{array}{c}
  L_{00} \future{  U_{01}} \\ \whline
  L_{10} \future{ U_{01} } + \future{  L_{11}} \future{ U_{11}} \\ \hline
  L_{20} \future{  U_{01}} + \future{ L_{21}} \future{ U_{11} }
  \end{array}
  \right)
  \wedge
  \left\vert
  P( 
  \left( \begin{array}{c}
  p_0 \\ \whline
  \future{  p_1} \\ \hline
  \future{  p_2 }\\  
  \end{array}
  \right)
  )^{-1}
   \left( \begin{array}{c}
  0 \\
  \whline
  \future{ L_{11}} \\ \hline
  \future{ L_{21}}
  \end{array}
  \right)
  \right\vert
  \leq
  \left( \begin{array}{c}
  J \\ \whline
  J \\ \hline
  J
  \end{array}
  \right)
\]
so that
\[
\begin{array}{l}
P( p_0 )
   \left( \begin{array}{c}
  A_{01} \\ \whline
  A_{11} \\ \hline
  A_{21}
  \end{array}
  \right) 
  - 
  \left( \begin{array}{c}
  0 \\ \whline  \mypadding[-2pt]
  P( 
  \left( \begin{array}{c}
  \future{  p_1} \\ \hline
  \future{  p_2}
  \end{array}
  \right)
  )^{-1}
  \left( \begin{array}{c}
  L_{10} \\ \hline
  L_{20}
  \end{array}
  \right)
  \future{   U_{01}}
  \end{array} \right)
  =
   \left( \begin{array}{c}
  L_{00} \future{ U_{01}} \\ \whline  \mypadding[-2pt]
  P( 
  \left( \begin{array}{c}
  \future{  p_1}\\ \hline
  \future{  p_2}
  \end{array}
  \right)
  )^{-1}
  \left( \begin{array}{c}
  \future{ L_{11}} \\ \hline
  \future{ L_{21}}
  \end{array}
  \right) U_{11}
  \end{array}
  \right) 
\wedge
  \left\vert
  P( 
  \left( \begin{array}{c}
  \future{  p_1} \\ \hline
  \future{  p_2 }\\  
  \end{array}
  \right)
  )^{-1}
   \left( \begin{array}{c}
  \future{ L_{11}} \\ \hline
  \future{ L_{21}}
  \end{array}
  \right)
  \right\vert
  \leq
  \left( \begin{array}{c}
  J \\ \hline
  J
  \end{array}
  \right).
  \end{array}
\]
If we first update
\[
\begin{array}{l}
\left( \begin{array}{c}
  A_{01} \\ \whline
  A_{11} \\ \hline
  A_{21}
  \end{array}
  \right)  :=
  P( p_0 )
   \left( \begin{array}{c}
  A_{01} \\ \whline
  A_{11} \\ \hline
  A_{21}
  \end{array}
  \right) 
  \\
  A_{01} := L_{00}^{-1} A_{01} ~~\mbox{(triangular solve with multiple rhs's)} \\
  \left( \begin{array}{c}
  A_{11} \\ \hline
  A_{21}
  \end{array}
  \right)
  :=
  \left( \begin{array}{c}
  A_{11} \\ \hline
  A_{21}
  \end{array}
  \right)
  -
\left( \begin{array}{c}
  A_{10} \\ \hline
  A_{20}
  \end{array}
  \right)
  A_{01}
  \end{array}
\]
then the updated parts of $ A $ (denoted by $ A^{\mbox{\rm upd.}} $) satisfy
\begin{eqnarray*}
\left( \begin{array}{c}
  A_{01} \\ \whline  \mypadding[-2pt]
  \left( \begin{array}{c}
  A_{11} \\ \hline
  A_{21}
  \end{array}
  \right)
  \end{array} \right)
  =
  \left( \begin{array}{c}
  U_{01} \\ \whline \mypadding[-2pt]
  \left( \begin{array}{c}
  A_{11}^{\mbox{\rm upd.}} \\ \hline
  A_{21}^{\mbox{\rm upd.}}
  \end{array}
  \right)
  \end{array} \right)
&\wedge&
  \left( \begin{array}{c}
  A_{11}^{\mbox{\rm upd.}} \\ \hline
  A_{21}^{\mbox{\rm upd.}}
  \end{array}
  \right)
  =
  P( 
  \left( \begin{array}{c}
  \future{  p_1} \\ \hline
  \future{  p_2}
  \end{array}
  \right)
  )^{-1}
  \left( \begin{array}{c}
  \future{ L_{11} }\\ \hline
  \future{ L_{21}}
  \end{array}
  \right)
  \future{ U_{11}}
   \\
& 
\wedge  &
  \left\vert
  P( 
  \left( \begin{array}{c}
  \future{  p_1} \\ \hline
  \future{  p_2 }\\  
  \end{array}
  \right)
  )^{-1}
   \left( \begin{array}{c}
  \future{ L_{11}} \\ \hline
  \future{ L_{21}}
  \end{array}
  \right)
  \right\vert
  \leq
  \left( \begin{array}{c}
  J \\ \hline
  J
  \end{array}
  \right)
\end{eqnarray*}
which can be rewritten as
\begin{eqnarray*}
\left( \begin{array}{c}
  A_{01} \\ \whline \mypadding[-2pt]
  \left( \begin{array}{c}
  A_{11} \\ \hline
  A_{21}
  \end{array}
  \right)
  \end{array} \right)
  =
  \left( \begin{array}{c}
  U_{01} \\ \whline \mypadding[-2pt]
  \left( \begin{array}{c}
  A_{11}^{\mbox{\rm upd.}} \\ \hline
  A_{21}^{\mbox{\rm upd.}}
  \end{array}
  \right)
  \end{array} \right)
&\wedge&
P( \future{ p_1} )
  \left( \begin{array}{c}
  A_{11}^{\mbox{\rm upd.}} \\ \hline
  A_{21}^{\mbox{\rm upd.}}
  \end{array}
  \right)
  =
  \left( \begin{array}{c}
  \future{ L_{11} }\\ \hline
  P( 
  \future{  p_2}
  )^{-1}
  \future{ L_{21}}
  \end{array}
  \right)
  \future{ U_{11}}
   \\
& 
\wedge &
  \left\vert
   \left( \begin{array}{c}
  \future{ L_{11} }\\ \hline
  P( 
  \future{  p_2}
  )^{-1} \future{ L_{21}}
  \end{array}
  \right)
  \right\vert
  \leq
  \left( \begin{array}{c}
  J \\ \hline
  J
  \end{array}
  \right).
\end{eqnarray*}
Hence, by Definition~\ref{def:LUpiv}, we complete the change in state for the middle blocks of columns from (\ref{eqn:step6}) to 
(\ref{eqn:step7}) by executing
\[
[ \left( \begin{array}{c}
A_{11} \\ \hline
A_{21}
\end{array} \right), p_1 ]
:= 
\mbox{\sc LUpiv} (
\left( \begin{array}{c}
A_{11} \\ \hline
A_{21}
\end{array} \right) ).
\]
Finally, examining how the contents of $ 
\left( \begin{array}{c}
A_{10} \\ \hline
A_{20} 
\end{array} \right) $ must 
change from (\ref{eqn:step6}) to 
(\ref{eqn:step7})
tells us that we must update
\[
\left( \begin{array}{c}
A_{10} \\ \hline
A_{20} 
\end{array} \right)  :=
P( p_1 ) \left( \begin{array}{c}
A_{10} \\ \hline
A_{20} 
\end{array} \right) .
\]
The complete algorithm is given in Figure~\ref{fig:LUpiv-blk-all}, as are the algorithms that correspond to the other invariants.  

What is important here is that the updates were determined by a sequence of substitutions of equivalent expressions in the predicates and reasoning about what assignments will change the state of the variables as prescribed.  
In other words, we have exposed it to be systematic.
The other blocked algorithms can be derived similarly.

\begin{figure}[p]
\input LUpiv_blk_all

\begin{center}
  \FlaAlgorithm  
\end{center}

\caption{Blocked algorithms for LU factorization with partial pivoting ({\sc LUpiv}).}
\label{fig:LUpiv-blk-all}
\end{figure}

\section{Implementation considerations}

While correct algorithms are essential for computing {\sc LUpiv}, there are also practical considerations related to implementation.
We briefly discuss these here.

\subsection{Choosing from the family of algorithms}
\label{sec:choosing}

A question is which algorithm to pick from the derived family.  
{\sc LUpiv} is a frequently-used operation when solving a linear system of equations and is at the heart of the LINPACK benchmark~\cite{LINPACK_Benchmark}. The impact of choice of algorithm on performance has thus been well-studied. 
Here are some considerations:
\begin{itemize}
    \item 
Blocked algorithms cast most computation in terms of so-called matrix-matrix operations (level-3 BLAS)~\cite{BLAS3}, which amortize the cost of moving data between memory layers~\cite{LAPACK}.  
Thus, at the top level, one should choose a blocked algorithm if the matrix is reasonably large.
\item
Blocked Variants~3piv-a and~3piv-b cast most computation in terms of the triangular solve with multiple right-hand sides (TRSM) $ A_{01} := L_{00}^{-1} A_{01} $ and the matrix-panel multiplication 
\begin{equation}
    \label{eqn:matrix-panel}
\left( \begin{array} {c}
A_{11} \\ \hline
A_{21} 
\end{array} \right)
:=  
\left(
\begin{array} {c}
A_{11} \\ \hline
A_{21} 
\end{array} \right) - \left( \begin{array} {c}
A_{10} \\ \hline
A_{20} 
\end{array} \right) A_{12}  .
\end{equation}
Blocked Variant~4piv casts most computation in terms of 
(\ref{eqn:matrix-panel})
and
$
A_{12} := A_{12} - A_{10} A_{02} 
$
(panel-matrix multiplication).
TRSM, matrix-panel, and panel-matrix multiplications generally do not achieve the best performance on single and multicore processors~\cite{Goto1,BLIS1}.
In contrast, blocked Variant 5piv casts most computation in terms of $ A_{22} := A_{22} - A_{21} A_{12} $ (rank-k update), which is the shape that tends to achieve the best performance on single and multicore processors~\cite{Goto1}.
For this reason, blocked Variant~5piv is typically employed.
\item
Rank-k updates also parallelize well to distributed memory architectures, making blocked Variant~5piv the algorithm of choice there as well~\cite{JPDC.dense}.
\item
The blocked algorithm requires {\sc LUpiv} as a suboperation and multiple levels of blocking may be employed.  At each level, a choice can be made as as to which variant to select.  Once the matrix becomes ``tall and skinny,'' the choice of algorithmic variant may change to one other than Variant~5piv.  The reason is that the TRSM now involves a very small matrix, and hence its performance becomes insignificant.  Also, matrix-panel multiplication and panel-matrix multiplication incur roughly half the memory operations that a rank-k update incurs when the panel is thin.
\item
When matrix $ A $ becomes large enough that it doesn't fit in the aggregate memories of the processors, out-of-core (OOC) techniques are employed to compute with data stored on disk or some other slower level of memory.  
In this case, left-looking algorithms (Variant~3piv-a for LU factorization) are typically employed because they reduce the traffic between secondary memory and main memory~\cite{Scott1,ICPP95}.  Furthermore, it is easier to add checkpointing and restarting to an algorithm that has an invariant where submatrices are either completely updated or not yet modified, as is the case for Invariant~3piv-a.  
\end{itemize}
The point is: under different circumstances, different variants have merit.
A complete study of what {\sc LUpiv} algorithm to use when goes beyond the scope of this paper.  
In~\cite{phd:low}, it is more generally discussed how the relationship between the invariant and the PME allows one to deduce beneficial properties of the resulting algorithm.

\subsection{Representing algorithms in code}

\begin{figure}[p]
\input LUpiv_libflame
\caption{Blocked {\sc LUpiv} (Variant 5piv) as implemented in libflame.  It has been slightly edited to fit the page.  The complete listing is at~\cite{libflame_LUpiv}.}
\label{fig:LUpiv_libflame}
\end{figure}

Central to the FLAME methodology for deriving correct algorithms is the avoidance of detailed indexing into the matrices.  
If the resulting algorithms are then translated by hand into code that involves indices, the opportunity for introducing programming errors reappears.
Because of this, we introduced the FLAME APIs~\cite{FLAME:API}, which allow the code to closely reflect the algorithms presented in FLAME notation.
Such APIs exist for C (FLAMEC), Matlab/Octave (FLAME@lab), and Python (FlamePy).  
In Figure~\ref{fig:LUpiv_libflame} we give Variant 5piv as implemented in libflame~\cite{libflame_github,libflame_LUpiv} with the FLAMEC API.
The parameter {\tt cntl} in that implementation passes what we call the {\em control tree} for the operation.  It is a hierarchical instruction that encodes what algorithmic variant to use in the various operations encountered in the body of the loop, including what variant to employ for the recursive {\sc LUpiv} that is invoked.  This allows one to compose a specialized implementation of {\sc LUpiv} for the different situations discussed in Section~\ref{sec:choosing}.

The FLAME APIs have their roots in an API,
PLAPACK~\cite{PLAPACK}, that was developed for coding DLA algorithms targeting distributed memory architectures.  A refinement and  more modern instantiation of the  ideas pioneered by PLAPACK is the C++ distributed-memory parallel library Elemental~\cite{Elemental:TOMS}, which also used a FLAME-like API to translate algorithms presented in FLAME notation to code.

\section{Conclusion}
\label{sec:conclusion}

The primary purpose of this paper was to demonstrate that the FLAME methodology can be employed to derive a family of loop-based algorithms for LU factorization with partial pivoting.  
All the classic algorithms  overwrite $ A $ so that upon completion 
$ A = L \backslash U \wedge P(p) \widehat A = L U \wedge \vert L \vert \leq J $.  
The algorithms derived from Invariant 3piv-b may not have been previously published and may have applicability when targeting distributed memory architectures,  if combining row swaps reduces communication overhead.
The fact that most of the derived algorithms were already known should not be a surprise: LU factorization with pivoting (Gaussian elimination with row swapping) is a very well-studied subject that goes back thousands of years~\cite{GRCAR2011163}.  Thus, it is encouraging that all known loop-based algorithms for this particular specification of the operation were found.
It remains to be worked out how to make systematic the dependence analysis that identifies potential loop invariants that support pivoting.

\NoShow{
\color{red}
Theorem provers leverage an algorithm that computes a result to prove theorems about that result. For example, proving that LU without pivoting completes if and only if all leading principle submatrices are nonsingular is proved in this way with the ACL2 theorem prover~\cite{10918400}.
Now that algorithms have been derived, the worksheet could be used as a framework to prove the well-known result that LUpiv completes (with an upper triangular matrix with nonzero diagonal) if and only if the columns of $ A $ are linearly independent.  In other words, it might be beneficial to incorporate FLAME-like techniques to automatic theorem provers.
}

\NoShow{
This paper raises some natural followup questions related to LU factorization with pivoting:
\begin{itemize}
\item 
How can one perform the dependence analyses needed to identify promising loop invariants from the PME and to properly sequence the assignments in the update? 
\item 
The specific formulation of LUpiv that we tackled is the one supported by, for example, LAPACK~\cite{LAPACK}.  Its predecessor, LINPACK~\cite{LINPACK},  performs pivoting slightly differently:  rows of $ L_{20} $ are not swapped.  The problem with this is that in the end solving $ A x = b $ becomes more difficult, since the pivoting of the right-hand side has to be intertwined with the lower triangular solve.
Can the FLAME methodology can be employed to derive algorithms for this alternative operation?
\end{itemize}
We leave these for further study.
}

Early incarnations of the techniques presented in this paper~\cite{LUpivArXiv} have enabled us to derive new (higher performing) algorithms for a less studied operation: the computation of the $ L T L^T $ factorization of a skew-symmetric matrix~\cite{LTLt}, with symmetric pivoting for stability.  While it can be argued that no exciting new algorithms resulted from this paper, the novel techniques have broad and impactful applicability beyond the well-understood setting of LU with pivoting.


\section*{Acknowledgements}
The FLAME methodology is the result of collaborations with many of our colleagues at UT Austin and around the world, most of whom appear as authors of works cited in this paper.  

This material is based upon work supported by the Correctness for Scientific Computing Systems (CS2), a joint program of the National Science Foundation (NSF) and the U.S. Department of Energy (DOE),
Office of Science, Office of Advanced Scientific Computing Research,
under NSF Award Numbers  
CCF-2446143 (CMU), CCF-2446145 (UT Austin), and DOE Award Number DE-SC0025950 (SMU).

{\em Any opinions, findings, and conclusions or recommendations expressed in this material are those of the author(s) and do not necessarily reflect the views of the National Science Foundation or the Department of Energy.
}


\bibliography{biblio}

@string{TOMS = "ACM Trans. Math. Soft."}

@string{JPDC = "J. Parallel Distrib. Comput."}

@string{FEB = "Feb."}

@string{MAR = "March"}

@string{APR = "April"}

@string{MAY = "May"}

@string{JUN = "June"}

@string{JUL = "July"}

@string{SEP = "Sept."}

@string{OCT = "Oct."}

@string{NOV = "Nov."}

@string{DEC = "Dec."}

@BOOK{LINPACK,
AUTHOR = {J. J. Dongarra and J. R. Bunch and C. B. Moler and G. W. Stewart},
TITLE = {LINPACK Users' Guide},
PUBLISHER = {SIAM},
ADDRESS = {Philadelphia},
YEAR = 1979 }

@BOOK{GVL4,
AUTHOR = {Gene H. Golub and Charles F. Van Loan},
TITLE = {Matrix Computations},
PUBLISHER = {The Johns Hopkins University Press},
ADDRESS = {Baltimore},
EDITION = {4nd},
YEAR = 2013,
doi ={https://doi.org/10.1137/1.9781421407944.}}

@ARTICLE{BLAS3,
OPTAUTHOR = {Jack J. Dongarra and Du Croz, Jeremy and Sven Hammarling and
Iain Duff},
AUTHOR = {Jack Dongarra and others},
TITLE = {A Set of Level 3 Basic Linear Algebra Subprograms},
JOURNAL = TOMS,
OPTVOLUME = 16,
OPTNUMBER = 1,
OPTPAGES = {1-17},
OPTMONTH = MAR,
YEAR = 1990,
doi = {https://doi.org/10.1145/77626.79170}}

@Article{JPDC.dense,
  author = 	 {Jack Dongarra and Robert {v}an~{d}e~{G}eijn and David Walker},
  title = 	 {Scalability Issues Affecting the Design of 
    a Dense Linear Algebra Library},
  journal = 	 JPDC,
  year = 	 {1994},
  OPTkey = 	 {},
  volume = 	 {22},
  number = 	 {3},
  month = 	 SEP,
  OPTpages = 	 {},
  OPTnote = 	 {},
  OPTannote = 	 {},
  doi = {https://doi.org/10.1006/jpdc.1994.1108}
}

@InProceedings{ICPP95,
  author = 	 {Ken Klimkowski and Robert {v}an~{d}e~{G}eijn},
  title = 	 {Anatomy of an out-of-core
dense linear solver},
  booktitle = 	 {Proceedings of the {I}nternational {C}onference on {P}arallel {P}rocessing 1995},
  OPTcrossref =  {},
  OPTkey = 	 {},
  OPTeditor = 	 {},
  volume = 	 {III - {A}lgorithms and {A}pplications},
  OPTnumber = 	 {},
  OPTseries = 	 {},
  year = 	 {1995},
  OPTorganization = {},
  OPTpublisher = {},
  OPTaddress = 	 {},
  OPTmonth = 	 {},
  pages = 	 {29--33},
  OPTnote = 	 {},
  OPTannote = 	 {},
  doi = {https://doi.org/10.1145/1055531.1055534}
}

@Book{LAPACK,
  author = 	 {E. Anderson and Z. Bai and J. Demmel and
                  J. E. Dongarra and  J. DuCroz and A. Greenbaum and
S. Hammarling and A. E. McKenney and S. Ostrouchov and D. Sorensen},
  title = 	 {{LAPACK} Users' Guide},
  publisher = 	 {SIAM},
  year = 	 {1992},
  OPTkey = 	 {},
  OPTeditor = 	 {},
  OPTvolume = 	 {},
  OPTnumber = 	 {},
  OPTseries = 	 {},
  address = 	 {Philadelphia},
  OPTedition = 	 {},
  OPTmonth = 	 {},
  OPTnote = 	 {},
  OPTannote = 	 {},
  doi={https://doi.org/10.1137/1.9780898719604}
}

@Book{PLAPACK,
  author = 	 {Robert A. {v}an~{d}e~{G}eijn},
  title = 	 {Using {PLAPACK}: Parallel Linear Algebra Package},
  publisher = 	 {The MIT Press},
  year = 	 {1997},
  OPTkey = 	 {},
  OPTeditor = 	 {},
  OPTvolume = 	 {},
  OPTnumber = 	 {},
  OPTseries = 	 {},
  OPTaddress = 	 {},
  OPTedition = 	 {},
  OPTmonth = 	 {},
  OPTnote = 	 {},
  OPTannote = 	 {},
  url = "https://mitpress.mit.edu/9780262720267/using-plapack-parallel-linear-algebra-package/"
}

@string{Feb = "February"}

@string{Mar = "March"}

@string{Apr = "April"}

@string{May = "May"}

@string{Jun = "June"}

@string{Jul = "July"}

@string{Sep = "September"}

@string{Oct = "October"}

@string{Nov = "November"}

@string{Dec = "December"}

@string{February = "February"}

@string{June = "June"}

@string{September = "September"}

@InProceedings{Scott1,
  author = 	 {David S. Scott},
  title = 	 {Out of core dense solvers on {I}ntel parallel supercomputers},
  booktitle = 	 {Proceedings of the {F}ourth {S}ymposium on the {F}rontiers of {M}assively {P}arallel {C}omputation},
  year = 	 {1992},
  pages = 	 {484--487},
  OPTkey = 	 {},
  OPTeditor = 	 {},
  OPTvolume = 	 {},
  OPTnumber = 	 {},
  OPTseries = 	 {},
  OPTaddress = 	 {},
  OPTmonth = 	 {},
  OPTorganization = {},
  OPTpublisher = {},
  OPTnote = 	 {},
  OPTannote = 	 {},
  doi = 
"10.1109/FMPC.1992.234876"
}

@Book{Gries,
  author = 	 {David Gries},
  title = 	 {The Science of Programming},
  publisher = 	 {Springer-Verlag},
  year = 	 {1981},
  OPTkey = 	 {},
  OPTvolume = 	 {},
  OPTnumber = 	 {},
  OPTseries = 	 {},
  OPTaddress = 	 {},
  OPTedition = 	 {},
  OPTmonth = 	 {},
  OPTnote = 	 {},
  OPTannote = 	 {},
  doi={https://doi.org/10.1007/978-1-4612-5983-1}
}

@InProceedings{FLAME_WoCo,
  author = 	 {John A. Gunnels and Robert A. {v}an~{d}e~{G}eijn},
  title = 	 {Formal Methods for High-Performance Linear 
                  Algebra Libraries},
  booktitle = 	 {The Architecture of Scientific Software},
  OPTcrossref =  {},
  OPTkey = 	 {},
  pages = 	 {193--210},
  year = 	 {2001},
  editor = 	 {Ronald F. Boisvert and Ping Tak Peter Tang},
  OPTvolume = 	 {},
  OPTnumber = 	 {},
  OPTseries = 	 {},
  OPTaddress = 	 {},
  OPTmonth = 	 {},
  OPTorganization = {},
  publisher = {Kluwer Academic Press},
  OPTnote = 	 {},
  OPTannote = 	 {},
  doi={https://doi.org/10.1007/978-0-387-35407-1_12}
}

@PhdThesis{Gunnels:PhD,
  author = 	 {John~A. Gunnels},
  title = 	 {A Systematic Approach to the Design and Analysis of Parallel Dense Linear Algebra Algorithms},
  school = 	 {Department of Computer Sciences,
                  The University of Texas},
  year = 	 {2001},
  OPTkey = 	 {},
  OPTtype = 	 {},
  OPTaddress = 	 {},
  month = 	 DEC,
  OPTnote = 	 {},
  OPTannote = 	 {},
  url={https://www.cs.utexas.edu/users/flame/pubs/FLAWN6.pdf}
}

@Article{Recipe,
  author =       "Paolo Bientinesi and John A. Gunnels and Margaret E. Myers and 
                 Enrique S. Quintana-Ort\'{\i} and Robert A. {v}an~{d}e~{G}eijn",
  title =        "The Science of Deriving Dense Linear Algebra Algorithms",
  journal =      TOMS,
  volume =       "31",
  number =       "1",
  month =        mar,
  year =         "2005",
  pages =        "1--26",
  URL =          "http://doi.acm.org/10.1145/1055531.1055532",
}

@Article{FLAME,
  author =       "John A. Gunnels and Fred G. Gustavson and Greg M. Henry and Robert A. {v}an~{d}e~{G}eijn",
  title =        "{FLAME}: {F}ormal {L}inear {A}lgebra {M}ethods {E}nvironment",
  journal =      TOMS,
  volume =       "27",
  number =       "4",
  month =        DEC,
  year =         "2001",
  pages =        "422--455",
  URL =          "http://doi.acm.org/10.1145/504210.504213", 
}

@Book{Dijkstra,
  author = 	 {Esger~W.~Dijkstra},
  ALTeditor = 	 {},
  title = 	 {A discipline of programming},
  publisher = 	 {Prentice Hall},
  year = 	 {1976},
  OPTkey = 	 {},
  OPTvolume = 	 {},
  OPTnumber = 	 {},
  OPTseries = 	 {},
  OPTaddress = 	 {},
  OPTedition = 	 {},
  OPTmonth = 	 {},
  OPTnote = 	 {},
  OPTannote = 	 {},
  doi = {https://dl.acm.org/doi/book/10.5555/550359}
}

@Book{Higham:2002:ASN,
  author =       "Nicholas J. Higham",
  title =        "Accuracy and Stability of Numerical Algorithms",
  publisher =    "Society for Industrial and Applied Mathematics",
  address =      "Philadelphia, PA, USA",
  edition =      "Second",
  year =         "2002",
  pages =        "xxx+680",
  ISBN =         "0-89871-521-0",
  doi="https://doi.org/10.1137/1.9780898718027"
}

@Article{Goto1,
author = {Goto, Kazushige and {v}an~{d}e~{G}eijn, Robert A.},
title = {Anatomy of high-performance matrix multiplication},
year = {2008},
issue_date = {May 2008},
publisher = {Association for Computing Machinery},
address = {New York, NY, USA},
volume = {34},
number = {3},
issn = {0098-3500},
url = {https://doi.org/10.1145/1356052.1356053},
doi = {10.1145/1356052.1356053},
journal = {ACM Trans. Math. Softw.},
month = may,
articleno = {12},
numpages = {25}
}

@Article{FLAME:API,
  author =       "Paolo Bientinesi and Enrique S. Quintana-Ort\'{\i} and Robert A. 
                 {v}an~{d}e~{G}eijn",
  title =        "Representing Linear Algebra Algorithms in Code: The {FLAME} 
                  Application Programming Interfaces",
  journal =      TOMS,
  volume =       "31",
  number =       "1",
  month =        mar,
  year =         "2005",
  pages =         "27--59",
  URL =          "http://doi.acm.org/10.1145/1055531.1055533",
}

@Article{Quintana-Orti:2003:FDA,
  author =       "Enrique S. Quintana-Ort\'{\i} and Robert A. {v}an~{d}e~{G}eijn",
  title =        "Formal Derivation of Algorithms: The Triangular {Sylvester} Equation",
  journal =      TOMS,
  volume =       "29",
  number =       "2",
  pages =        "218--243",
  month =        jun,
  year =         "2003",
  CODEN =        "ACMSCU",
  ISSN =         "0098-3500",
  doi =          "http://doi.acm.org/10.1145/779359.779365",
}

@PhdThesis{Paolo:PhD,
  author = 	 {Paolo Bientinesi},
  title = 	 {Mechanical Derivation and Systematic Analysis of Correct Linear Algebra Algorithms},
  school = 	 {Department of Computer Sciences,
                  The University of Texas},
  year = 	 {2006},
  note = 	 {Technical Report  TR-06-46. September 2006},
  url={http://hdl.handle.net/2152/2679}
}

@Article{Bientinesi:2008:FAR,
  author =       "Paolo Bientinesi and Brian Gunter and Robert A. {v}an~de~{G}eijn",
  title =        "Families of Algorithms Related to the Inversion of a Symmetric Positive Definite Matrix",
  journal =      TOMS,
  volume =       "35",
  number =       "1",
  month =        jul,
  year =         "2008",
  pages =        "3:1--3:22",
  URL =          "http://doi.acm.org/10.1145/1377603.1377606"
}

@Book{TSoPMC,
  author =       {Robert A. {v}an~{d}e~{G}eijn and Enrique S. Quintana-Ort\'{\i}},
  ALTeditor =    {},
  title =        {The Science of Programming Matrix Computations},
  url =    {www.lulu.com},
  year =         {2008}
}

@Book{libflame_ref,
  author =       {Field G. {V}an~{Z}ee},
  ALTeditor =    {},
  title =        {{\tt libflame}: {T}he {C}omplete {R}eference},
url =    {www.lulu.com},
  year =         {2009}
}

@article{Elemental:TOMS,
 author = {Poulson, Jack and Marker, Bryan and {v}an~{d}e~{G}eijn, Robert A. and Hammond, Jeff R. and Romero, Nichols A.},
 OPTauthor = {Poulson, Jack and others},
 title = {Elemental: A New Framework for Distributed Memory Dense Matrix Computations},
 journal = {ACM Trans. Math. Softw.},
 OPTissue_date = {February 2013},
 OPTvolume = {39},
 OPTnumber = {2},
 OPTmonth = feb,
 year = {2013},
 OPTissn = {0098-3500},
 OPTpages = {13:1--13:24},
 OPTarticleno = {13},
 OPTnumpages = {24},
 OPTurl = {http://doi.acm.org/10.1145/2427023.2427030},
 doi = {10.1145/2427023.2427030},
 OPTacmid = {2427030},
 OPTpublisher = {ACM},
 OPTaddress = {New York, NY, USA},
}

@Article{CiSE09,
  author =   {Field G. {V}an~{Z}ee and  Ernie Chan and Robert van~de~Geijn and Enrique S. Quintana-Ort\'{\i} and Gregorio Quintana-Ort\'{\i}},
  title =    {The libflame Library for Dense Matrix Computations},
 journal =   {{IEEE} Computation in Science \& Engineering},
 year =      {2009},
 volume =    {11},
 number =    {6},
 pages =     {56--62},
 doi = "https://dl.acm.org/doi/abs/10.5555/2220074.2220181"
}

@article{BLIS1,
 author = {Van Zee, Field G. and {v}an~{d}e~{G}eijn, Robert A.},
 OPTauthor = {Van Zee, Field and others},
 title = {{BLIS}: A Framework for Rapidly Instantiating {BLAS} Functionality},
 journal = {ACM Trans. Math. Softw.},
 OPTissue_date = {June 2015},
 OPTvolume = {41},
 OPTnumber = {3},
 OPTmonth = jun,
 year = {2015},
 OPTissn = {0098-3500},
 OPTpages = {14:1--14:33},
 OPTarticleno = {14},
 OPTnumpages = {33},
 OPTurl = {http://doi.acm.org/10.1145/2764454},
 doi = {10.1145/2764454},
 OPTacmid = {2764454},
 OPTpublisher = {ACM},
 OPTaddress = {New York, NY, USA},
}

@article{Bientinesi:2011:GMS:2078718.2078728,
 author = {Bientinesi, Paolo and {v}an~{d}e~{G}eijn, Robert A.},
 title = {Goal-Oriented and Modular Stability Analysis},
 journal = {SIAM J. Matrix Anal. Appl.},
 issue_date = {February 2011},
 volume = {32},
 number = {1},
 month = mar,
 year = {2011},
 issn = {0895-4798},
 pages = {286--308},
 numpages = {23},
 url = {http://dx.doi.org/10.1137/080741057},
 doi = {10.1137/080741057},
 acmid = {2078728},
 publisher = {Society for Industrial and Applied Mathematics},
 address = {Philadelphia, PA, USA},
}

@phdthesis{phd:low,
    author = "Tze~Meng Low",
    title  = "A Calculus of Loop Invariants for Dense Linear Algebra Optimization",
    school = "The University of Texas at Austin, Department of Computer Science",
    year   = 2013,
    month  = dec,
    URL = "https://www.cs.utexas.edu/~flame/pubs/low_dissertation_20139.pdf"
}

@article{EWD:EWD340, author = {Dijkstra, Edsger W.}, title = {The humble programmer}, year = {1972}, issue_date = {Oct. 1972}, publisher = {Association for Computing Machinery}, address = {New York, NY, USA}, volume = {15}, number = {10}, issn = {0001-0782}, url = {https://doi.org/10.1145/355604.361591}, doi = {10.1145/355604.361591}, journal = {Commun. ACM}, month = oct, pages = {859–866}, numpages = {8} }

@article{Eijkhout20101805,
title = "Towards mechanical derivation of Krylov solver libraries",
journal = "Procedia Computer Science",
volume = "1",
number = "1",
pages = "1805 - 1813",
year = "2010",
note = "ICCS 2010",
issn = "1877-0509",
doi = "10.1016/j.procs.2010.04.202",
url = "http://www.sciencedirect.com/science/article/pii/S1877050910002036",
author = "Victor Eijkhout and Paolo Bientinesi and Robert {v}an~{d}e~{G}eijn"
}

@Article{Bientinesi2013,
author="Bientinesi, Paolo
and Gunnels, John A.
and Myers, Margaret E.
and Quintana-Ort{\'i}, Enrique S.
and Rhodes, Tyler
and {v}an~{d}e~{G}eijn, Robert A.
and {V}an~{Z}ee, Field G.",
title="Deriving dense linear algebra libraries",
journal="Formal Aspects of Computing",
year="2013",
month="Nov",
day="01",
volume="25",
number="6",
pages="933--945",
issn="1433-299X",
doi="10.1007/s00165-011-0221-4",
url="https://doi.org/10.1007/s00165-011-0221-4"
}

@article{10.1145/363235.363259,
author = {Hoare, C. A. R.},
title = {An Axiomatic Basis for Computer Programming},
year = {1969},
issue_date = {Oct. 1969},
publisher = {Association for Computing Machinery},
address = {New York, NY, USA},
volume = {12},
number = {10},
issn = {0001-0782},
url = {https://doi.org/10.1145/363235.363259},
doi = {10.1145/363235.363259},
journal = {Commun. ACM},
month = {oct},
pages = {576–580},
numpages = {5}
}

@phdthesis{Cl1ck,
   author = "Diego Fabregat-Traver",
   title  = " Knowledge-based automatic generation of linear algebra algorithms and code",
   school = "RWTH Aachen",
   year   = 2014,
   month  = apr,
   url    = "http://arxiv.org/abs/1404.3406"
}

@inproceedings{TC_Correctness,
	author = {Lee, Matthew and Low, Tze Meng},
	title = {A Family of Provably Correct Algorithms for Exact Triangle Counting},
	year = {2017},
	isbn = {9781450351270},
	publisher = {Association for Computing Machinery},
	address = {New York, NY, USA},
	url = {https://doi.org/10.1145/3145344.3145484},
	doi = {10.1145/3145344.3145484},
	booktitle = {Proceedings of the First International Workshop on Software Correctness for HPC Applications},
	pages = {14–20},
	numpages = {7},
	location = {Denver, CO, USA},
	series = {Correctness'17}
}

@inbook{10.1145/3544585.3544597,
author = {{v}an {d}e {G}eijn, Robert and Myers, Maggie},
title = {Applying Dijkstra’s Vision to Numerical Software},
year = {2022},
isbn = {9781450397735},
publisher = {Association for Computing Machinery},
address = {New York, NY, USA},
edition = {1},
doi = {https://doi.org/10.1145/3544585.3544597},
booktitle = {Edsger Wybe Dijkstra: His Life,Work, and Legacy},
pages = {215–230},
numpages = {16}
}

@INPROCEEDINGS{9835383,
  author={Acosta, Jay A. and Low, Tze Meng and Parikh, Devangi N.},
  booktitle={2022 IEEE International Parallel and Distributed Processing Symposium Workshops (IPDPSW)}, 
  title={Families of Butterfly Counting Algorithms for Bipartite Graphs}, 
  year={2022},
  volume={},
  number={},
  pages={304-313},
  doi={10.1109/IPDPSW55747.2022.00060}}

@book{10.1145/3544585, editor = {Apt, Krzysztof R. and Hoare, Tony}, title = {Edsger Wybe Dijkstra: His Life,Work, and Legacy}, year = {2022}, isbn = {9781450397735}, publisher = {Association for Computing Machinery}, address = {New York, NY, USA}, edition = {1}, volume = {45}, 
doi = {https://doi.org/10.1145/3544585} }

@book{LAFF-On-Correctness,
author = "Robert A. van de Geijn and Margaret E. Myers",
title = "LAFF-On Programming for Correctness",
publisher = "\url{ulaff.net}"
}

@article{LINPACK_Benchmark,
author = {Dongarra, Jack J. and Luszczek, Piotr and Petitet, Antoine},
title = {The LINPACK Benchmark: past, present and future},
journal = {Concurrency and Computation: Practice and Experience},
volume = {15},
number = {9},
pages = {803-820},
doi = {https://doi.org/10.1002/cpe.728},
OPTurl = {https://onlinelibrary.wiley.com/doi/abs/10.1002/cpe.728},
eprint = {https://onlinelibrary.wiley.com/doi/pdf/10.1002/cpe.728},
year = {2003}
}

@misc{libflame_LUpiv,
key="fla\_lu\_piv",
title="{\tt FLA\_LU\_piv\_blk\_var5}",
url="https://github.com/flame/libflame/blob/master/src/lapack/dec/lu/piv/vars/flamec/FLA_LU_piv_blk_var5.c",
year="2014"
}

@misc{libflame_github,
  key = {libflame},
  title = {libflame},
  year = {2023},
  publisher = {GitHub},
  journal = {GitHub repository},
  howpublished = {\url{https://github.com/flame/libflame}},
  commit="7e94bbd0cb203d7fbd43d30c1890614ceac3b02a"
}

@article{GRCAR2011163,
title = {How ordinary elimination became Gaussian elimination},
journal = {Historia Mathematica},
volume = {38},
number = {2},
pages = {163-218},
year = {2011},
issn = {0315-0860},
doi = {https://doi.org/10.1016/j.hm.2010.06.003},
url = {https://www.sciencedirect.com/science/article/pii/S0315086010000376},
author = {Joseph F. Grcar}
}

@INPROCEEDINGS{10918400,
  author={Kwan, Carl and Hunt, Warren A.},
  booktitle={2024 Formal Methods in Computer-Aided Design (FMCAD)}, 
  title={Automatic Verification of Right-Greedy Numerical Linear Algebra Algorithms}, 
  year={2024},
  volume={},
  number={},
  pages={242-250},
  doi={10.34727/2024/isbn.978-3-85448-065-5_30}}

@misc{LTLt,
      title={Performant Tridiagonal Factorization of Skew-Symmetric Matrices}, 
      author={Ishna Satyarth and Chao Yin and Devin A. Matthews and Maggie Myers and Robert van de Geijn and RuQing G. Xu},
      year={2026},
      eprint={2411.09859},
      archivePrefix={arXiv},
      primaryClass={cs.MS},
      note={To appear in the SIAM Journal on Scientific Computing},
      url={https://arxiv.org/abs/2411.09859}, 
}

@InProceedings{kwan_et_al:LIPIcs.ITP.2024.25,
  author =	{Kwan, Carl and Hunt Jr., Warren A.},
  title =	{{Formalizing the Cholesky Factorization Theorem}},
  booktitle =	{15th International Conference on Interactive Theorem Proving (ITP 2024)},
  pages =	{25:1--25:16},
  series =	{Leibniz International Proceedings in Informatics (LIPIcs)},
  ISBN =	{978-3-95977-337-9},
  ISSN =	{1868-8969},
  year =	{2024},
  volume =	{309},
  editor =	{Bertot, Yves and Kutsia, Temur and Norrish, Michael},
  publisher =	{Schloss Dagstuhl -- Leibniz-Zentrum f{\"u}r Informatik},
  address =	{Dagstuhl, Germany},
  URL =		{https://drops.dagstuhl.de/entities/document/10.4230/LIPIcs.ITP.2024.25},
  URN =		{urn:nbn:de:0030-drops-207532},
  doi =		{10.4230/LIPIcs.ITP.2024.25}
}

@misc{LUpivArXiv,
      title={Formal Derivation of LU Factorization with Pivoting}, 
      author={Robert van de Geijn and Maggie Myers},
      year={2023},
      eprint={2304.03068v1},
      archivePrefix={arXiv},
      primaryClass={cs.MS},
      url={https://arxiv.org/abs/2304.03068v1}, 
      doi = "https://doi.org/10.48550/arXiv.2304.03068v1"
}

\appendix



\section{Proof of Theorem~\ref{thm:iamax}}
\label{appendix:proof}

\begin{proof}
    We need to show the following assertion is {\em  true}:
    \[
\begin{array}{l}
      \left\{ P:
    \left( \begin{array}{c}
    \alpha_1 \\ \hline
    a_2
    \end{array} \right)
    =
    \left( \begin{array}{c}
    \widehat \alpha_1 \\ \hline
    \widehat a_2
    \end{array} \right)
    \wedge
    P( \future{\pi}) 
    \left( \begin{array}{c}
    \widehat \alpha_1 \\ \hline
    \widehat a_2
    \end{array} \right)
    =
    \left( \begin{array}{c}
    1 \\ \hline
    \future{l_2}
    \end{array} \right)
    \future{\upsilon_1}
    \wedge
    \left\vert
    \left( \begin{array}{c}
    1 \\ \hline
    \future{ l_2}
    \end{array} \right)
    \right\vert
    \leq
    \left( \begin{array}{c}
    1 \\ \hline
    j
    \end{array} \right)
    \right\} \\ \mypadding[-2pt]
    S_0: \pi := \mbox{\sc iamax}( \left( \begin{array}{c}
    \alpha_1 \\ \hline
    a_2
    \end{array} \right)  ) \\ \mypadding[-2pt]
    S_1: \left( \begin{array}{c}
    \alpha_1 \\ \hline
    a_2
    \end{array} \right) 
    := P( \pi )
    \left( \begin{array}{c}
    \alpha_1 \\ \hline
    a_2
    \end{array} \right) \\ \mypadding[-2pt]
    S_2: a_2 := a_2 / \alpha_1 \\ \mypadding[-2pt]
        \left\{ Q:
    \left( \begin{array}{c}
    \alpha_1 \\ \hline
    a_2
    \end{array} \right)
    =
    \left( \begin{array}{c}
    \upsilon_1 \\ \hline
    l_2
    \end{array} \right)
    \wedge
    P( \pi) 
    \left( \begin{array}{c}
    \widehat \alpha_1 \\ \hline
    \widehat a_2
    \end{array} \right)
    =
    \left( \begin{array}{c}
    1 \\ \hline
    l_2
    \end{array} \right)
    \upsilon_1
    \wedge
    \left\vert
    \left( \begin{array}{c}
    1 \\ \hline
    l_2
    \end{array} \right)
    \right\vert
    \leq
    \left( \begin{array}{c}
    1 \\ \hline
    j
    \end{array} \right)
    \right\}
    \end{array}
    \]
where $ P $ and $ Q $ are the precondition and postcondition for {\sc LUpiv}, respectively, for the case where $ A = \left(  \begin{array}{c}
\alpha_1 \\ \hline
a_2 
\end{array}
\right) $, which implicitly means that $ \upsilon_1 $, a $ 1 \times 1 $ upper-triangular matrix) is nonzero.

    Our proof strategy is as follows:
To prove 
$\{P\} S_0; S_1; S_2 \{Q\}$ is {\em true}, one must prove
that the following assertions hold:
{\renewcommand{\arraystretch}{1.1}
\[
\begin{array}{l}
    \{ P \} \\
    S_0 \\
    \{ P_1: \text{wp}( ``S_1", P_2 ) \} \\
    S_1 \\
    \{ P_2: \text{wp}( ``S_2", Q  )\} \\
    S_2 \\
    \{ Q \}
\end{array}
\]}
We do so by computing $ P_2 $ and $ P_1 $ (in that order), and then reasoning that the definition of {\sc iamax} means that 
$ \{ P \} S_0 \{ P_1 \} $  is {\em true}.

Thus
\[
\begin{array}{l}
      \left\{ P:
    \left( \begin{array}{c}
    \alpha_1 \\ \hline
    a_2
    \end{array} \right)
    =
    \left( \begin{array}{c}
    \widehat \alpha_1 \\ \hline
    \widehat a_2
    \end{array} \right)
    \wedge
    P( \future{\pi}) 
    \left( \begin{array}{c}
    \widehat \alpha_1 \\ \hline
    \widehat a_2
    \end{array} \right)
    =
    \left( \begin{array}{c}
    1 \\ \hline
    \future{l_2}
    \end{array} \right)
    \future{\upsilon_1}
    \wedge
    \left\vert
    \left( \begin{array}{c}
    1 \\ \hline
    \future{ l_2}
    \end{array} \right)
    \right\vert
    \leq
    \left( \begin{array}{c}
    1 \\ \hline
    j
    \end{array} \right)
    \right\} \\ \mypadding[-2pt]
    S_0: \pi := \mbox{\sc iamax}( \left( \begin{array}{c}
    \alpha_1 \\ \hline
    a_2
    \end{array} \right)  ) \\ \mypadding[-2pt]
        \left\{
    \begin{array}{l@{}c@{}l}
    P_1: \mbox{wp}(``\left( \begin{array}{c}
    \alpha_1 \\ \hline
    a_2
    \end{array} \right) 
    := P( \pi )
    \left( \begin{array}{c}
    \alpha_1 \\ \hline
    a_2
    \end{array} \right)", P_2 ) \\ \mypadding[-2pt]
    ~~~~~ =
    \left[
    P( \pi )
\left( \begin{array}{c}
    \alpha_1 \\ \hline
    a_2 
    \end{array} \right)
    =
    \left( \begin{array}{c}
    1 \\ \hline
    l_2  
    \end{array} \right) \upsilon_1
    \wedge
    P( \pi) 
    \left( \begin{array}{c}
    \widehat \alpha_1 \\ \hline
    \widehat a_2
    \end{array} \right)
    =
    \left( \begin{array}{c}
    1 \\ \hline
    l_2
    \end{array} \right)
    \upsilon_1
    \wedge
    \left\vert
    \left( \begin{array}{c}
    1 \\ \hline
    l_2
    \end{array} \right)
    \right\vert
    \leq
    \left( \begin{array}{c}
    1 \\ \hline
    j
    \end{array} \right)
    \right]  \\ \mypadding[-2pt]
    ~~~~~ =
    \left[
\left( \begin{array}{c}
    \alpha_1 \\ \hline
    a_2 
    \end{array} \right)
    =
    \left( \begin{array}{c}
    \widehat \alpha_1 \\ \hline
    \widehat a_2 
    \end{array} \right)
    \wedge
    P( \pi) 
    \left( \begin{array}{c}
    \widehat \alpha_1 \\ \hline
    \widehat a_2
    \end{array} \right)
    =
    \left( \begin{array}{c}
    1 \\ \hline
    l_2
    \end{array} \right)
    \upsilon_1
    \wedge
    \left\vert
    \left( \begin{array}{c}
    1 \\ \hline
    l_2
    \end{array} \right)
    \right\vert
    \leq
    \left( \begin{array}{c}
    1 \\ \hline
    j
    \end{array} \right)
    \right]
    \end{array} 
    \right\}
    \\ \mypadding[-2pt]
    S_1: \left( \begin{array}{c}
    \alpha_1 \\ \hline
    a_2
    \end{array} \right) 
    := P( \pi )
    \left( \begin{array}{c}
    \alpha_1 \\ \hline
    a_2
    \end{array} \right) \\ \mypadding[-2pt]
    \left\{
    \begin{array}{l@{}c@{}l}
    P_2: \mbox{wp}(``a_2 := a_2 / \alpha_1", Q )  \\ \mypadding[-2pt]
    ~~~~~ =
    \left[
\left( \begin{array}{c}
    \alpha_1 \\ \hline
    a_2 / \alpha_1
    \end{array} \right)
    =
    \left( \begin{array}{c}
    \upsilon_1 \\ \hline
    l_2
    \end{array} \right)
    \wedge
    P( \pi) 
    \left( \begin{array}{c}
    \widehat \alpha_1 \\ \hline
    \widehat a_2
    \end{array} \right)
    =
    \left( \begin{array}{c}
    1 \\ \hline
    l_2
    \end{array} \right)
    \upsilon_1
    \wedge
    \left\vert
    \left( \begin{array}{c}
    1 \\ \hline
    l_2
    \end{array} \right)
    \right\vert
    \leq
    \left( \begin{array}{c}
    1 \\ \hline
    j
    \end{array} \right)
    \right] \\ \mypadding[-2pt]
    ~~~~~ =
    \left[
\left( \begin{array}{c}
    \alpha_1 \\ \hline
    a_2 
    \end{array} \right)
    =
    \left( \begin{array}{c}
    1 \\ \hline
    l_2  
    \end{array} \right) \upsilon_1
    \wedge
    P( \pi) 
    \left( \begin{array}{c}
    \widehat \alpha_1 \\ \hline
    \widehat a_2
    \end{array} \right)
    =
    \left( \begin{array}{c}
    1 \\ \hline
    l_2
    \end{array} \right)
    \upsilon_1
    \wedge
    \left\vert
    \left( \begin{array}{c}
    1 \\ \hline
    l_2
    \end{array} \right)
    \right\vert
    \leq
    \left( \begin{array}{c}
    1 \\ \hline
    j
    \end{array} \right)
    \right]
    \end{array}
    \right\}
    \\ \mypadding[-2pt]
    S_2: a_2 := a_2 / \alpha_1 \\ \mypadding[-2pt]
        \left\{ Q:
    \left( \begin{array}{c}
    \alpha_1 \\ \hline
    a_2
    \end{array} \right)
    =
    \left( \begin{array}{c}
    \upsilon_1 \\ \hline
    l_2
    \end{array} \right)
    \wedge
    P( \pi) 
    \left( \begin{array}{c}
    \widehat \alpha_1 \\ \hline
    \widehat a_2
    \end{array} \right)
    =
    \left( \begin{array}{c}
    1 \\ \hline
    l_2
    \end{array} \right)
    \upsilon_1
    \wedge
    \left\vert
    \left( \begin{array}{c}
    1 \\ \hline
    l_2
    \end{array} \right)
    \right\vert
    \leq
    \left( \begin{array}{c}
    1 \\ \hline
    j
    \end{array} \right)
    \right\}
    \end{array}
    \]
    This leaves us to reason that
    \[
\begin{array}{l}
      \left\{ P:
    \left( \begin{array}{c}
    \alpha_1 \\ \hline
    a_2
    \end{array} \right)
    =
    \left( \begin{array}{c}
    \widehat \alpha_1 \\ \hline
    \widehat a_2
    \end{array} \right)
    \wedge
    P( \future{ \pi}) 
    \left( \begin{array}{c}
    \widehat \alpha_1 \\ \hline
    \widehat a_2
    \end{array} \right)
    =
    \left( \begin{array}{c}
    1 \\ \hline
    \future{l_2}
    \end{array} \right)
    \future{\upsilon_1}
    \wedge
    \left\vert
    \left( \begin{array}{c}
    1 \\ \hline
    \future{ l_2}
    \end{array} \right)
    \right\vert
    \leq
    \left( \begin{array}{c}
    1 \\ \hline
    j
    \end{array} \right)
    \right\} \\ \mypadding[-2pt]
    S_0: \pi := \mbox{\sc iamax}( \left( \begin{array}{c}
    \alpha_1 \\ \hline
    a_2
    \end{array} \right)  ) \\ \mypadding[-2pt]
        \left\{
    P_1:  =
\left( \begin{array}{c}
    \alpha_1 \\ \hline
    a_2 
    \end{array} \right)
    =
    \left( \begin{array}{c}
    \widehat \alpha_1 \\ \hline
    \widehat a_2 
    \end{array} \right)
    \wedge
    P( \pi) 
    \left( \begin{array}{c}
    \widehat \alpha_1 \\ \hline
    \widehat a_2
    \end{array} \right)
    =
    \left( \begin{array}{c}
    1 \\ \hline
    l_2
    \end{array} \right)
    \upsilon_1
    \wedge
    \left\vert
    \left( \begin{array}{c}
    1 \\ \hline
    l_2
    \end{array} \right)
    \right\vert
    \leq
    \left( \begin{array}{c}
    1 \\ \hline
    j
    \end{array} \right)
    \right\}
    \end{array}
\]
Here $ P $ says that there exists an index $ \pi $ such  that permutation brings the largest element in magnitude (known to  be nonzero) to the top, etc.  The command  {\sc iamax} computes that index, and hence the Haore triple evaluates  to {\em true}.
\end{proof}

\end{document}